\documentclass[iop,jphysa,reprint,article-title]{revtex4-1} % options: preprint or reprint, draft, linenumbers, ...
\usepackage[utf8]{inputenc} % use UTF-8 encoding in bibliography

\usepackage{cmap} % make ligatures etc. copy- and searchable
\usepackage[T1]{fontenc} % needed to have copy-able umlaute
\usepackage{lmodern}
\usepackage[scr]{rsfso} % calligraphic math font Formal Script

\usepackage{amsmath,amssymb,amsfonts}
\usepackage{mathtools,mleftright}
\usepackage{graphicx}
\graphicspath{{cherryfigs/}}
\newlength\figwidth
\usepackage[inline]{enumitem}

\usepackage{url,hyperref}
\usepackage[usenames,dvipsnames]{xcolor}
\hypersetup{colorlinks=true, linkcolor=BrickRed, urlcolor=blue!50!black, citecolor=blue!50!black}

\usepackage[capitalise]{cleveref}
\crefname{equations}{Eqs.}{Eqs.}
\Crefname{equations}{Equations}{Equations}
\creflabelformat{equations}{\textup{(#2#1#3)}}

\renewcommand{\vec}[1]{\mathbf{#1}}
\newcommand\diff{\mathrm{d}}
\newcommand\e{\text{e}}
\renewcommand\i{\text{i}}
\DeclareMathOperator\Real{Re}

\renewcommand\geq\geqslant
\renewcommand\leq\leqslant
\renewcommand\epsilon\varepsilon

\newcommand{\FPT}{\text{FPT}}
\newcommand{\In}{\texttt{in}}
\newcommand{\Out}{\texttt{out}}

\begin{document}

%opening
\title{Generalized master equation for first-passage problems in partitioned spaces}

\author{Daniela Frömberg}
\email{daniela.froemberg@fu-berlin.de}
\affiliation{
  Freie Universität Berlin,
  Fachbereich Mathematik und Informatik,
  Arnimallee 6,
  14195 Berlin, Germany
}

\author{Felix Höf{}ling}
\email{f.hoefling@fu-berlin.de}
\affiliation{
  Freie Universität Berlin,
  Fachbereich Mathematik und Informatik,
  Arnimallee 6,
  14195 Berlin, Germany
}
\affiliation{
  Zuse Institute Berlin,
  Takustr. 7,
  14195 Berlin, Germany
}

\date{\today}

\begin{abstract}
Motivated by a range of biological applications related to the transport of molecules in cells,
we present a modular framework to treat first-passage problems for diffusion in partitioned spaces.
The spatial domains can differ with respect to their diffusivity, geometry, and dimensionality, but can also refer to transport modes alternating between diffusive, driven, or anomalous motion.
The approach relies on a coarse-graining of the motion by dissecting the trajectories on domain boundaries or when the mode of transport changes, yielding a small set of states.
The time evolution of the reduced model follows a generalized master equation (GME) for non-Markovian jump processes; the GME takes the form of a set of linear integro-differential equations in the occupation probabilities of the states and the corresponding probability fluxes.
Further building blocks of the model are partial first-passage time (FPT) densities, which encode the transport behavior in each domain or state.
After an outline of the general framework for multiple domains,
the approach is exemplified and validated for a target search problem with two domains in one- and three-dimensional space, first by exactly reproducing known results for an artificially divided, homogeneous space, and second by considering the situation of domains with distinct diffusivities.
Analytical solutions for the FPT densities are given in Laplace domain and are complemented by numerical backtransforms yielding FPT densities over
many decades in time, confirming that the geometry and heterogeneity of the space can introduce additional characteristic time scales.
\end{abstract}

\maketitle

\section{Introduction}

% diffusion in partitioned spaces

Transport within heterogeneous media is ubiquitous in nature, and finds rich expression
in biological contexts.
Cellular spaces and membranes show heterogeneous structures due to compartmentalization and macromolecular crowding, leading to
complex diffusive transport of molecules with implications for biochemical reactions \cite{Zhou:2008,Hoefling:2013,Weiss:2014}.
The phenomenon of anomalous (sub-)diffusion plays certainly a prominent role here, which is typically more pronounced for large molecules and long-distance transport and which is likely to subsume a number of physical causes.
Yet, already concrete microscopic structures can give rise to non-trivial dynamics,
examples being spatial domains of varying diffusivity \cite{Dross:PLoS2009},
possibly separated by the nuclear envelope or other diffusion barriers \cite{Schlimpert:Cell2012},
and the nowadays established nano-scale partitioning of the plasma membrane
\cite{Sezgin:Cell2015, Raghupathy:Cell2015, Koldso:JPCB2016}.
Another aspect are alternating modes of motion, e.g., stochastic switching between actively directed and Brownian motion \cite{Witzel:BJ2019}, or between different dimensionalities of space
such as sliding on one-dimensional (1D) DNA strands and 3D diffusion in the nucleoplasm \cite{Mirny:JPA2009};
in the context of nanocatalysts, one finds surface diffusion on a nanoparticle interleaved with 3D diffusion in solution \cite{Lin:2020}.

In addition, numerous technological and physical applications rely on the peculiar transport properties in multi-phase materials, with examples ranging from hydrogen storage \cite{Li:2012}
over microfluidic devices and molecular sieving \cite{Han:2008, MolecularSieves:2005, Cerbelli:MNFl2013}
to flows in geological sediments and porous media \cite{Sahimi:1993, Dentz:GeoRL09}.
In contrast to random media,
we have situations in mind where the medium is composed of few elementary building blocks (domains), which may occur repeatedly.
A typical goal in studies of heterogeneous materials is homogeneization, that is to obtain an effective description of the macroscopic transport by coarse graining the problem up to spatial scales at which the medium can be regarded as homogeneous (see Refs.~\cite{Bur:PRE17, Dentz:GeoRL09, Spanner:PRL2016, Adrover:PhFl2019} for examples).
Different to this, the present work aims at retaining the heterogeneous character of the medium, while keeping only
statistical information on the transport in each domain.
This allows one to capture both long-range transport (e.g., effective diffusivities) and local behaviour (e.g., return probabilities) within the same model.

For a variety of applications, the transport on the single-trajectory level is relevant, with diffusion-influenced chemical reactions as a prominent situation. A specific example are enzyme cascades, where spatial proximity can induce a channeling of the substrate molecules \cite{Kuzmak:SR2019}.
In this and related situations, the arrival of the first molecules matters more than the behaviour of the bulk, and trajectory-resolved statistics are more meaningful than the mean values \cite{Mattos:PRE2012, Grebenkov:CC2018}.
Of particular interest are first-passage times (FPTs), which measure the time of first encounter between substrate molecules diffusing in space.
The reciprocal of the mean FPT is essentially the reaction rate constant in classical reaction kinetics, based on the law of mass action, which is applicable only for high copy numbers of the reactants and under well-mixed conditions.
The influence of a geometric confinement on the FPT distributions and thus the reaction kinetics was appreciated only recently \cite{Mattos:PRE2012, Benichou:2014, Grebenkov:CC2018, Grebenkov:NJP2019}.
Already for comparably simple geometries, one observes FPT distributions that are governed by two or more widely distinct time scales, meaning that the FPTs can fluctuate considerably from one molecular trajectory to the other
\cite{Benichou:2014, Godec:SR2016}.

\begin{figure}
\includegraphics{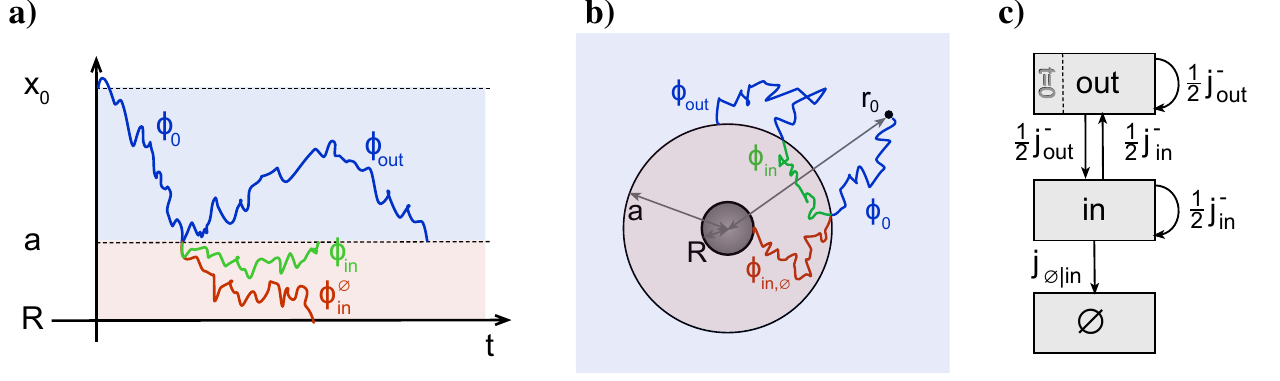}%
\caption{Geometries of the first-passage problem with two domains and sketches of exemplary partial trajectories
  in a one-dimensional half axis (panel~a) and between concentric spheres in three dimensions (b);
  the boundary of the larger sphere with radius $b > |\vec r_0|$ is not shown.
  Space is partitioned at the position/radius $a$ into an inner (red) and an outer (blue) domain.
  The symbols $\phi$ denote the dwell time densities of the different types of partial trajectories.
  ~(c) Graph of the simplified two-domain model (without an auxiliary shell). Arrows indicate the transitions between the states and are annotated by the corresponding fraction of the loss fluxes $j^-$ from each state.
  }
  \label{fig:ChPn}
\end{figure}

The aim of the present paper is a mathematical framework of first-passage problems for diffusion in a continuous space partitioned into domains.
These domains are assumed to be homogeneous regions, which however can differ from each other with respect to their transport properties or even their dimensionality; at a later stage, different chemical reactions may occur within each domain.
We do explicitly not require any mechanism of barrier crossing between the domains, although the approach allows for such barriers.
A computationally motivated example for a situation without barriers is Doi's volume reaction model \cite{Doi:1975a}, used in particle-based reaction--diffusion simulations \cite{Erban:2009,Dibak:JCP2019}.
In the presence of sufficiently high barriers, each domain forms a metastable set with respect to a molecule's motion, transport on long time scales resembles a Markovian hopping process, and the reaction kinetics in such a partitioned space can be described by a spatio-temporal chemical master equation \cite{Winkelmann:JCP2016}. Here, we lay a basis to go beyond these approximations.

Exploiting the Markov property of (idealized) diffusion, we coarse-grain the process and
dissect a given trajectory into parts whenever a domain boundary is crossed (\cref{fig:ChPn}a,b).
Following the dynamics from one crossing to the next, we have recast the diffusion problem as a renewal process.
These \emph{partial} trajectories are fully contained within a single domain, and all possible trajectories within one domain are identified with a coarse-grained state.
% We assign a certain state to the diffusing molecule depending on its current position and the corresponding partial trajectory.
%
Therewith, the original diffusion process in continuous space has been replaced by a continuous-time random walk (CTRW) on the domain states (\cref{fig:ChPn}c).
The waiting times between the jumps correspond to the dwell times (or residence times) on the respective partial trajectories
and their distributions are given as solutions to (partial) first-passage problems on each domain.
The jumps between domains constitute in general not a Poisson process (as for a Markovian random walk), not even for simple diffusion, and thus introduce a memory into the evolution equations of the occupation probabilities of the states.
We will refer to the latter set of equations as the \emph{generalized master equation} (GME) of the coarse-grained problem.
While this procedure is readily justified in case of a Markov process, a less stringent requirement is a renewal property of
the process at the domain boundaries, that is, the evolution in one domain needs to be
independent from the history in another domain.

Early studies of CTRWs on a finite state space date back almost half a century ago
% das war kurz bevor ich geboren bin *heul*
and include a two-state model for the orientational motion of molecules in dense media \cite{Lindenberg:1975},
later amended by a CTRW in space on top of it \cite{Weiss:1976};
a more recent application are the on and off times in blinking quantum dots \cite{Margolin:JPCB2006}.
In all these examples, the sought quantities were computed using classical renewal equations;
introductory texts on this method can be found in Refs.~\cite{Feller:ProbabilityBd2, Hughes:Random_Walks}.
In the present paper, we will adopt an alternative approach put forward by Chechkin \emph{et al.}~\cite{Chechkin:JPA2005},
where the authors also coined the term ``generalized master equation'' (GME).
This approach was applied successfully in the modeling of reaction-subdiffusion systems \cite{Froemberg:PRL2008, Froemberg:PRE2011}.
The charme of the method lies in its clarity as it can be derived from a few
basic principles such as local balance and continuity of probability fluxes.
Below, we will first outline the general framework for multiple domains, which can encompass rather complex and convoluted situations.
Noteworthy, the dwell times on a domain are permitted to depend on the part of the boundary through which a trajectory enters and exits, i.e., on the previous and following states, and therefore our treatment extends the lattice models studied so far.
For the detailed solutions, we will then adhere to the paradigmatic case of two domains, either with completely the same physical properties (i.e., no inhomogeneity at all) in order to establish the method (\cref{sec:1d,sec:rad}), or with two different diffusion constants (\cref{sec:application}).
Both cases allow for the comparison to known results: a textbook solution in the first case and recent literature in the second \cite{Godec:SR2016}, where the total FPT density was obtained by solving a partial-differential equation (PDE) on the whole, non-uniform domain.
In these examples, the symmetries of the setups allow for relatively straightforward, explicit calculations of the partial FPT densities within homogeneous domains.

\section{Generalized master equation for first-passage problems with multiple domains}

\subsection{Formulation of the problem in case of two domains}
\label{sec:problem}

Before introducing the general framework, we formulate the problem for the case of two domains amended by an absorbing target, which serves as an illustration and a test bed of the proposed GME to calculate the FPT distribution for target search.
We consider free diffusion
\emph{(i)}~on a 1D half axis with an absorbing boundary at position $R > 0$, and
\emph{(ii)}~in the 3D space between two concentric spheres of radii $R < b$, where the boundary of the inner sphere is absorbing and the outer one is reflecting (\cref{fig:ChPn}a,b).
A domain boundary is placed at the position $a > R$ (1D) and at radius $a$ with $b > a > R$ (3D), respectively, and we refer to the region between $R$ and $a$ as inner domain and to the remaining accessible space as outer domain.
The trajectories start in the outer domain at $\vec x_0$ with $|\vec x_0| > a$.
(For simplicity, we use the vector notation also in 1D space.)
The central quantity of interest is the probability density $p_\FPT(t)$ of the random time $t$ until the first encounter with the boundary at $R$.

The trajectories are dissected into partial ones whenever the boundary at $|\vec x| = a$ is hit.
By the Markov property of diffusion, the evolution of a partial trajectory inside a domain depends on the entry point at the boundary, but not on the motion in the other domain. The partial trajectory ends when reaching another point of the domain boundary.
The dwell time on a given partial trajectory (equivalently, in its respective domain) is the time the molecule spends between two consecutive events of hitting the domain boundary, which include entrance to and exit from the domain, but also merely touching the boundary.
The probability density of dwell times is an FPT density itself, which is why it will be referred to as a \emph{partial} FPT density in the following.
In the outer domain, there are two types of trajectories: starting at $\vec x_0$ and starting at the boundary $|\vec x| = a$, both types end at this boundary; the corresponding partial FPT densities are $\phi_0(t)$ and $\phi_\Out(t)$.
Trajectories in the inner domain always start at $|\vec x| = a$, but either leave the domain at this boundary again or are stopped at $|\vec x| = R$ with partial FPT densities $\phi_\In(t)$ and $\phi_{\In}^{\emptyset}(t)$, respectively.
The specific forms of these dwell time densities for the geometries chosen here are discussed in \cref{sec:partFPT1d,sec:partFPTrad}.

As a first test, we assume no physical difference between the inner and outer domains so that the boundary at $|\vec x| = a$ is only an artificial one. This allows us to check the proposed GME approach of dissecting trajectories and assembling the overall FPT density $\phi_\FPT(t)$ for reaching the boundary $|\vec x| = R$ from the partial ones.
The result must reproduce the known distribution of FPTs for reaching $|\vec x|=R$ directly when starting at $\vec x_0$.
In \cref{sec:application}, we elaborate on the situation of different diffusion coefficients in the two domains.

\subsection{Generalized master equation}
\label{ssec:GME}

The derivation of the GME of the present problem follows ideas in Ref.~\cite{Chechkin:JPA2005}
and makes use of two types of conservation laws:
\begin{enumerate*}
 \item probability is conserved locally: net gains or losses within one state determine the change of local probability, and

 \item probability is conserved in the transitions between the states (continuity of the fluxes).
\end{enumerate*}
The other key ingredient is an equation accounting for the renewal character of the stochastic process, 
linking present losses from a state with the gains at a previous time.

For the general scheme, we consider a partition of the accessible space into $n$ domains labeled by $\alpha \in \{1,\dots,n\}$.
The trajectory of a diffusing molecule is dissected whenever it hits a boundary between domains, and the label of the corresponding domain is assigned to each partial trajectory.
This gives rise to a set of coarse-grained states and to the probabilities $\rho_\alpha(t)$ that the molecule is found in domain $\alpha$ at time~$t$.
If two domains $\alpha$ and $\beta$ share a boundary $\alpha|\beta$, then transitions across this boundary induce a probability flux.
Accounting for the direction of the transition, $j_{\alpha|\beta}(t)$ denotes the flux from state $\beta$ to state $\alpha$ at time $t$.
Transitions from a domain to itself are possible ($j_{\alpha|\alpha} \neq 0$) since at the boundary of a domain the origin of the trajectory is irrelevant by the assumed renewal property.
The overall probability gain of state $\alpha$ is $j_\alpha^+ = \sum_\beta j_{\alpha|\beta}$,
whereas the total loss flux reads
$j_\alpha^- = \sum_\beta j_{\beta|\alpha}$. Local conservation of probability [principle (i)] then implies
\begin{equation}
 \frac{\diff}{\diff t} \rho_\alpha(t) = j_\alpha^+(t) - j_\alpha^-(t)  = \sum_\beta [j_{\alpha|\beta}(t) - j_{\beta|\alpha}(t)]\,,
 \label{eq:genlocalbal}
\end{equation}
that is, the temporal change of probability in a state is the difference between the total gain and loss fluxes.
Summing over $\alpha$ shows that the overall probability is conserved globally, as it should:
$(\diff/\diff t) \sum_\alpha \rho_\alpha(t) = 0$.

The continuity of the fluxes, principle (ii), determines the total gain fluxes as a fixed linear combination of the total losses,
\begin{equation}
	j_\alpha^+ = \sum_{\beta} w_{\alpha|\beta} j_\beta^- \,,
	\label{eq:genfluxbal}
\end{equation}
where the weights $w_{\alpha|\beta}$ encode the connectivity of the domains and form a stochastic $n\times n$ matrix with columns adding up to unity,
$\sum_{\alpha} w_{\alpha|\beta} = 1$.
The sum in \cref{eq:genfluxbal} is actually restricted to those domains $\beta$ that are adjacent to $\alpha$, otherwise we can put $w_{\alpha|\beta}=0$.
Each summand represents the partial gain of state $\alpha$ stemming from a loss of probability in state $\beta$. Thus, the probability flux across the boundary $\alpha|\beta$ directed towards $\alpha$ is given by the product
\begin{equation}
	j_{\alpha|\beta} = w_{\alpha|\beta} j_\beta^- .
	\label{eq:fluxcomponents}
\end{equation}
The $w_{\alpha|\beta}$ include the transmission probability $q_{\alpha|\beta}$ that a partial trajectory ending at the $\alpha|\beta$ boundary is indeed continued in the domain $\alpha$.
The fraction of the loss flux $j_\beta^-$ that reaches the boundary $\alpha|\beta$ is $w_{\alpha|\beta}/q_{\alpha|\beta}$ and sums to
unity, $\sum_{\alpha\neq\beta} w_{\alpha|\beta}/q_{\alpha|\beta} = 1$.
The flux $j_{\beta|\beta} = w_{\beta|\beta} j_\beta^-$ describes those trajectories that hit the boundary of domain $\beta$, but return and are continued inside of $\beta$.
If the transition probabilities at all boundaries of $\beta$ are equal, $q_{\alpha|\beta} =: q_\beta$, the fraction of such ``remainers'' is
\begin{equation}
  w_{\beta|\beta} = 1 - \sum_{\alpha\neq\beta} w_{\alpha|\beta} = 1 - q_\beta \,.
\end{equation}
For unbiased diffusion, $q_{\alpha|\beta} = 1/2$ at all boundaries $\alpha|\beta$, so that $w_{\beta|\beta} = 1/2$ is the probability of a trajectory to remain in the domain after reaching its boundary.
In the presence of one (or more) absorbing domain $\emptyset$, exits from such a domain cannot occur, $j_{\alpha|\emptyset} = 0$ and so $j_\emptyset^-=0$. The temporal change of $\rho_\emptyset(t)$ is the sought FPT density of the target search problem, $p_\FPT(t) = \diff\rho_\emptyset(t)/\diff t = j_\emptyset^+(t)$.

Invoking the picture of an ensemble of particles, a loss of particles from a domain at time $t$ can only happen
if the particles had been there before, either from the very beginning or through gains at an earlier time $t-\tau$,
where $\tau$ is the dwell time of a specific particle (i.e., a partial trajectory) in that domain.
The loss fluxes are thus linked to the gain fluxes and obey renewal-like relations \cite{Chechkin:JPA2005}, which for a domain $\beta$ reads schematically:
\begin{equation}
  j_\beta^-(t) = \int_0^t \phi_\beta(\tau) \, j_\beta^+(t - \tau)\,\diff \tau + \phi_\beta^{(0)}(t) \rho_\beta^{(0)}
  \label{eq:genlossflren}
\end{equation}
with $\phi_\beta(\tau)$ denoting the probability density of dwell times. The last term describes particles that were initially placed inside of the domain with probability $\rho_\beta^{(0)} = \rho_\beta(t=0)$ and reach the domain boundary at time $t$ according to the FPT density $\phi_\beta^{(0)}(t)$.
\Cref{eq:genlossflren} applies only under certain conditions, e.g., for a highly symmetric domain, as shown in \cref{sec:simplifiedGME}.

More generally, we distinguish the parts of the boundary of domain $\beta$ according to their adjacent domain and
consider the ``ports'' $\alpha|\beta$ that allow for transitions to and from a domain $\alpha$.
Then, the dwell time of a partial trajectory in $\beta$ starting at the boundary $\beta|\gamma$ and stopping at $\alpha|\beta$ is statistically characterized by its FPT density $\phi_{\beta|\gamma}^\alpha(t)$, with $\alpha$ and $\gamma$ taken from the set of adjacent domains.
The probability of leaving the domain $\beta$ through the boundary $\alpha|\beta$ after starting at $\beta|\gamma$ is referred to as splitting probability,
\begin{equation}
 P_{\beta|\gamma}^\alpha = \int_0^\infty \phi_{\beta|\gamma}^\alpha(t)\,\diff t \,;
\end{equation}
the splitting probabilities of the same initial boundary sum up to unity, $\sum_{\alpha\neq \beta} P_{\beta|\gamma}^\alpha  = 1$.

For the directed probability fluxes $j_{\alpha|\beta}$ through these ports, we postulate the following renewal equation as a generalization of \cref{eq:genlossflren}:
\begin{equation}
 j_{\alpha|\beta} = q_{\alpha|\beta}\sum_{\gamma\neq\beta} \phi_{\beta|\gamma}^\alpha *
  \mleft[j_{\beta|\gamma} + (q_{\gamma|\beta}^{-1} - 1) j_{\gamma|\beta}\mright]
  + q_{\alpha|\beta} \,\phi_{\beta|0}^\alpha \,\rho_\beta^{(0)} \,; \quad \alpha \neq \beta,
 \label{eq:genflren}
\end{equation}
omitting the time arguments and introducing the convolution symbol as
$
  ({f * g})(t) = \int_0^t f(\tau) \,g(t-\tau) \,\diff \tau
$
for integrable functions $f,g$.
In the last term of \cref{eq:genflren}, $\phi_{\beta|0}^\alpha(t)$ is the FPT density of trajectories starting initially in the interior of domain $\beta$ with probability $\rho_\beta^{(0)} := \rho_\beta(0)$ and hitting the boundary $\alpha|\beta$ at time $t$;
the initial position is either fixed at some point $\vec r_0$ within the domain, or $\phi_{\beta|0}^\alpha(t)$ is an average over a distribution of initial positions.
The multiplication of the r.h.s.\ of the renewal equation by the transmission probability $q_{\alpha|\beta}$ takes into account that the flux $j_{\alpha|\beta}$ includes only those trajectories that are continued in the domain $\alpha$, but the FPT densities describe only the arrival at the boundary $\alpha|\beta$.
Concerning the start point of the partial trajectories, there are two possibilities to reach the boundary $\beta|\gamma$, which are expressed by the brackets inside of the convolution:
(i)~entering the domain $\beta$ by a transition from $\gamma$ and
(ii)~touching the boundary from inside of $\beta$, i.e., remainers that were in $\beta$ before time $t-\tau$.
Whereas the flux corresponding to (i) is simply $j_{\beta|\gamma}(t-\tau)$, the second situation requires that the flux $j_{\gamma|\beta}(t-\tau)/q_{\gamma|\beta}$ of trajectories that reached the boundary from inside is multiplied by the probability $1-q_{\gamma|\beta}$ of not leaving to domain $\gamma$;
for unbiased diffusion, $(q_{\gamma|\beta}^{-1} - 1) = 1$.
The flux $j_{\beta|\beta} = w_{\beta|\beta} j_\beta^-$ due to self-transition is proportional to the overall flux to the neighbouring domains, where the prefactor follows with $j_\beta^- = \sum_\alpha j_{\alpha|\beta}$, so that
\begin{equation}
  j_{\beta|\beta} = \frac{w_{\beta|\beta}}{1 - w_{\beta|\beta}} \sum_{\alpha \neq \beta} j_{\alpha|\beta} \,;
  \label{eq:genselfflux}
\end{equation}
the prefactor is unity if $w_{\beta|\beta} = 1/2$ as for unbiased diffusion.

The renewal equation \cref{eq:genflren} simplifies considerably if, due to symmetries of the domain $\beta$, all boundaries are equivalent with respect to the partial FPT densities;
examples are a slab-like domain delimited by two parallel, infinite planes, and a domain that forms a regular simplex.
Then, there are only two types of FPT densities,
namely $\phi_\beta^{(d)} = \phi_{\beta|\gamma}^\alpha$ for trajectories connecting distinct boundaries and
$\phi_\beta^{(s)} = \phi_{\beta|\alpha}^\alpha$ for self-transitions ($\alpha=\gamma$).
Summation of \cref{eq:genflren} over the adjacent domains $\alpha$ of $\beta$ yields  \cref{eq:genlossflren} for the total loss flux $j_\beta^-$ with the FPT density for reaching some part of the boundary of $\beta$:
\begin{equation}
  \phi_\beta = \phi_\beta^{(s)} + (z_\beta - 1) \phi_\beta^{(d)} \,,
\end{equation}
where the coordination number $z_\beta$ counts the adjacent domains of $\beta$ (see \cref{sec:simplifiedGME}).
Further, the loss flux is equally distributed over all boundaries, i.e., $w_{\alpha|\beta} = q_\beta / z_\beta$ for $\alpha\neq\beta$ and $w_{\beta|\beta} = 1 - q_\beta$.

The set of linear eqs.~\eqref{eq:genlossflren} together with \cref{eq:genfluxbal} for $\beta=1,\dots,n$
can be solved for the $n$ loss fluxes $j_\beta^-$, provided that the weight matrix $(w_{\alpha|\beta})$ is known a priori; the FPT densities are considered model parameters.
Alternatively, \cref{eq:genflren,eq:genselfflux} specify a set of $2m$ linear equations that can be solved for the $2m$ directed fluxes $j_{\alpha|\beta}$, assuming that the system contains $m$ boundaries (ports) between domains.

\subsection{GME for the two-domain model with absorption}\label{ssec:2dom}

In the two-domain model of \cref{sec:problem}, the state of the molecule is described by the probabilities $\rho_\In(t)$ and $\rho_\Out(t)$ that at time $t$ it is found in the inner and outer domain, respectively.
It is amended by the probability $\rho_\emptyset(t)$ that the trajectory was stopped at time $t$ or earlier:
the molecule has reached the target and was removed from the system, e.g., by a chemical reaction.
The transitions between the states induce probability fluxes $j_{\alpha|\beta}$ with $\alpha,\beta \in \{\In, \Out\}$,
which lead to the total gain $(+)$ and loss $(-)$ fluxes, $j_{\In}^{\pm}$ and $j_{\Out}^{\pm}$, of the $\In$ and $\Out$ states, respectively.
In addition, there is a flux $j_{\emptyset|\In}$ indicating the trajectories that traversed the $\In$ state before being stopped at the target ($\emptyset$). In the subsequent notation, the flux $j_{\emptyset|\In}$ is treated as an additional loss from the $\In$ state and is \emph{not} included in the symbol $j_\In^-$.
\Cref{fig:ChPn}c shows a transition graph of the model along with the loss fluxes.

For an unbiased diffusion, a partial trajectory ending at the domain boundary $|\vec x| = a$ is continued in either the inner or the outer domain with equal probability $1/2$.
By the Markov property of diffusion, this applies for partial trajectories irrespective of which side of the boundary they belong to ($\In$ or $\Out$ state).
Thus, $j_{\In|\Out} = j_{\Out|\Out} = \frac{1}{2} j_\Out^-$, and we will not distinguish between $j_{\In|\Out}$ and $j_{\Out|\Out}$ in the following; correspondingly, $j_{\In|\In} = j_{\Out|\In} = \frac{1}{2} j_\In^-$.
Therefore, continuity of the fluxes in the transitions requires that [see \cref{eq:genfluxbal}]
\begin{equation}
  j_\Out^+(t) = \frac{1}{2} \left[j_\In^-(t) + j_\Out^-(t)\right]
  \qquad\text{and}\qquad
  j_\In^+(t) = \frac{1}{2} \left[j_\Out^-(t) + j_\In^-(t)\right],
  \label{flbalinout1}
\end{equation}
where the terms on the right represent the incoming arrows of the state given on the left as depicted in \cref{fig:ChPn}c;
it follows that $j_\Out^+(t) = j_\In^+(t)$.
Next, local balance in each state demands [\cref{eq:genlocalbal}]:
\begin{subequations}
\begin{align}
\frac{\diff}{\diff t} \rho_\Out(t) &= j_\Out^+(t) -j_\Out^-(t), \label{locbalout1} \\
\frac{\diff}{\diff t} \rho_\In(t) &= j_\In^+(t) - j_\In^-(t) - j_{\emptyset|\In}(t), \label{locbalin1} \\
\frac{\diff}{\diff t}\rho_{\emptyset}(t) &= j_{\emptyset|\In}(t). \label{FPpdfnaive}
\end{align}
\end{subequations}
One confirms readily that the overall probability, including that of the absorbed particles, is conserved:
$(\diff/\diff t)\left[ \rho_\In(t) + \rho_\Out(t) + \rho_{\emptyset}(t)\right] = 0$.
The quantity $\rho_\emptyset(t)$ in \cref{FPpdfnaive} collects the trajectories that have stopped up to time~$t$, and its
change is thus equal to the sought FPT density of the target search problem,
$\diff \rho_\emptyset(t) / \diff t = p_\FPT(t)$. So the task is to compute the flux $j_{\emptyset|\In}(t)$ onto the boundary $|\vec x| = R$.

Finally, the loss fluxes make a recursion to the gain fluxes and obey a set of renewal-like relations:
\begin{subequations}
	\begin{align}
	j_\Out^- (t) &= \phi_0(t) + \int_0^t \phi_\Out(t-t') \, j^+_\Out(t') \, \diff t' , \label{flrenout1}\\
	j_\In^-(t) &= \int_0^t \phi_\In(t-t') \, j^+_\In(t') \, \diff t'  , \label{flrenin1} \\
	j_{\emptyset|\In}(t) &= \int_0^t \phi_{\In}^{\emptyset}(t-t') \, j^+_\In(t') \, \diff t'  . \label{flreninstop1}
	\end{align}
\end{subequations}
These equations follow from \cref{eq:genlossflren}, which is applicable because in the two-domain problem merely two types of partial FPT densities occur:
the outer domain permits transitions only to the $\In$ state through the boundary at $|\vec x| = a$ and thus the only FPT density in the domain $\Out$ is $\phi_\Out := \phi_{\Out|\In}^\In$ for trajectories from this boundary to itself, along with the FPT density $\phi_0$ for the initial part of the trajectories starting at $\vec r_0$ in the interior of $\Out$.
Partial trajectories in the inner domain $\In$ can either end at the boundary $|\vec x| = a$ (flux $j_\In^-$) or at $|\vec x| = R$ (absorption, $j_{\emptyset|\In}$), see \cref{fig:ChPn}c.
Further, there are no gains to $\In$ from the absorbing boundary ($q_{\emptyset|\In}=1$), and
we have only the self-transition term corresponding to $\phi_\In = \phi_{\In|\Out}^\Out$ in \cref{flrenin1}.
The loss $j_{\emptyset|\In}$ to the absorbed state is a ``distinct'' contribution expressed by the FPT density $\phi_\In^\emptyset := \phi_{\In|\Out}^\emptyset$.
Note that $\phi_\In$ and $\phi_{\In}^{\emptyset}$ are no proper FPT densities in the sense that they do not normalize,
the full density of dwell times in the state $\In$ is given by their sum, which satisfies
$\int_0^\infty \mleft[\phi_\In(t) + \phi_{\In}^{\emptyset}(t) \mright] \,\diff t= 1$.
This fact expresses the splitting of probabilities to follow one or the other type of partial trajectory (i.e., to survive or to be stopped at the end of the current step).

The system of \cref{locbalout1,locbalin1,FPpdfnaive,flrenout1,flrenin1,flreninstop1}, together with \cref{flbalinout1}, forms the GME of the first-passage problem with two domains.
Using \cref{flbalinout1}, the gain fluxes $j_{\In}^+$ and $j_{\Out}^+$ can be expressed in terms of loss fluxes,
which leaves us with a system of 6 linear integro-differential equations in 6 variables (3 loss fluxes and 3 occupation probabilities). This type of equation system is conveniently solved in Laplace domain.

\subsection{Solution in Laplace domain and numerical backtransform}

The Laplace transform $\tilde{f} := \mathscr{L}[f]$ of a measurable function $f(t)$ on $\mathbb{R}_{\geq 0}$ is defined as \cite{Feller:ProbabilityBd2}
\begin{equation}
	\tilde{f}(u) = \mathscr{L}[f](u) := \int_0^\infty \e^{-ut} f(t) \,\diff t ,
	\label{LapTr}
\end{equation}
where the frequency $u > 0$ is the Laplace variable conjugate to~$t$.
For a convolution $(f * g)(t) = \int_0^t f(t-t') g(t') \diff t'$
it holds $\mathscr{L}[f * g] = \mathscr{L}[f] \mathscr{L}[g]$, and for the time derivative one has
$\mathscr{L}\mleft[\dot f\mright](u) = u\, \mathscr{L}[f](u) - f(0)$.

Laplace transformation of the self-consistent, linear system of integro-differential \cref{locbalout1,locbalin1,FPpdfnaive,flrenout1,flrenin1,flreninstop1},
yields a closed set of linear, algebraic equations in the probabilities and fluxes with the partial FPT densities as coefficients.
Substituting $j_{\In}^+$ and $j_{\Out}^+$ in \cref{flrenout1,flrenin1,flreninstop1} by means of \cref{locbalout1,locbalin1}, we obtain:
\begin{subequations}
	\begin{align}
	\tilde j_\Out^- (u) &= \tilde\phi_0(u) + \tilde\phi_\Out(u)
	\left[ u\tilde\rho_\Out(u) - 1 + \tilde j^-_\Out(u)\right] ,
	\label{joutL}\\
	\tilde j_\In^-(u) &= \tilde\phi_\In(u)
	\left[u \tilde\rho_\In(u) + \tilde j^-_\In(u) + \tilde j_{\emptyset|\In}(u)\right] ,
	\label{jinL}\\
	\tilde j_{\emptyset|\In}(u) &= \tilde\phi_{\In}^{\emptyset}(u)
	\left[u \tilde\rho_\In(u) + \tilde j^-_\In(u) + \tilde j_{\emptyset|\In}(u)\right] ,
	\label{jreacL}
	\end{align}
\end{subequations}
where we made use of the initial conditions $\rho_\Out(0) = 1$ and $\rho_\In(0) = \rho_{\emptyset}(0) = 0$.
Solving for the loss fluxes, we have:
\begin{subequations}
	\begin{align}
		\tilde j_\Out^- (u) &= \frac{\tilde \phi_0 (u) - \tilde\phi_\Out(u)}{1 - \tilde\phi_\Out(u)} +
		\frac{u \tilde\phi_\Out(u)}{1 - \tilde\phi_\Out(u)} \tilde\rho_\Out(u)
		\label{flrenout1L}\\
		\tilde j_\In^-(u) &=\frac{\tilde\phi_\In(u)}{1 - \tilde\phi_\In(u)}
		\left[u\tilde\rho_\In(u) + \tilde j_{\emptyset|\In}\right]
		\label{flrenin1L}\\
		\tilde j_{\emptyset|\In}(u) &=\frac{\tilde\phi_\In(u)}{1 - \tilde\phi_{\In}^{\emptyset}(u)}
		\left[u\tilde\rho_\In(u) + \tilde j^-_\In\right].
		\label{flreninstop1L}
	\end{align}	
\end{subequations}
The system of equations in Laplace domain is completed by
\begin{subequations}
\begin{align}
u \tilde\rho_\Out(u) - 1 &= \frac{1}{2}\left[\tilde j_\In^-(u) - \tilde j_\Out^-(u)\right]\label{urhooutnL}\\
u \tilde\rho_\In(u) &= \frac{1}{2}\left[\tilde j_\Out^-(u) - \tilde j_\In^-(u)\right] - \tilde j_{\emptyset|\In}(u) \label{urhoinnL}\\
u \tilde\rho_{\emptyset}(u) &= \tilde j_{\emptyset|\In}(u) \label{urhostopnL}
\end{align}
\end{subequations}
from \cref{locbalin1,locbalout1,FPpdfnaive}.
Solving the linear system for $\tilde j_{\emptyset|\In}(u)$ provides us with an explicit expression for the sought FPT density in Laplace domain, $\tilde p_{\text{FPT}}(u) = \tilde j_{\emptyset|\In}(u)$, which is fully specified by the partial FPT densities $\tilde\phi_x(u)$:
\begin{equation}
\tilde p_{\text{FPT}}(u) =
\frac{2 \tilde\phi_0(u) \tilde\phi_{\emptyset}(u)}{\tilde\phi_\In(u) \mleft[\tilde \phi_{\emptyset}(u) - 2\mright] - 2 \tilde\phi_\Out(u) + 4} \,;
\label{FPnaive}
\end{equation}
see Eqs.~\eqref{eq:sol_naive} of the appendix for the solution for all densities and fluxes.
As a by-product, our approach also provides the overall sojourn time in a certain state $\alpha \in \{\In, \Out\}$ up to a time $t$ simply by integrating $\rho_\alpha(t)$ over time, which amounts to calculating $\tilde\rho_\alpha(u) / u$ in the Laplace domain.

The actual FPT density $\phi_\FPT(t)$ in time domain can be obtained from a numerical Laplace backtransform.
The procedure is understood best by switching to the characteristic function of the FPTs, given by the one-sided Fourier transform $\hat \phi(\omega) = \int_0^\infty \e^{\i \omega t} \phi(t) \diff t$, which is well defined for all frequencies $\omega \in \mathbb{R}$ since $\phi(t)$ is a probability density and thus integrable.
It allows for the analytic continuation to the upper complex plane and is connected to the Laplace transform by $\tilde \phi(u) = \hat \phi(\i u)$.
The backtransform is uniquely given by a cosine transform:
\begin{equation}
 \phi(t) = \frac{2}{\pi} \int_0^\infty \! \cos(\omega t) \Real \hat \phi(\omega) \, \diff\omega \,,
\end{equation}
using that $\Real \hat \phi(\omega)$ is an even function in $\omega$.
For the robust numerical evaluation of the Fourier integral, we used a modified Filon quadrature as developed recently for the back and forth transformation between time correlation functions and their dissipation spectra \cite{Straube:CP2020}.

\subsection{Regularized GME with an auxiliary boundary}
\label{sec:GMEaux}

The coefficients in \cref{flrenout1L,flrenin1L,flreninstop1L} become singular if any of the FPT densities $\tilde\phi_x(u) = 1$, i.e., if $\phi_x(t) = \delta_+(t)$ is the density of the Dirac measure on the positive reals, $\mathbb{R}_{\geq 0}$, supported at $t=0$. Unfortunately, we are facing this problem for $\phi_\In(t)$ and $\phi_\Out(t)$ with the setup of \cref{fig:ChPn}:
partial trajectories starting at the boundary $|\vec x| = a$ return to that boundary immediately, almost surely, the starting and ending points are the same.
This issue is familiar from the first-return problem for diffusion in the continuum, it does not arise for random walks on a lattice.

\begin{figure}
	\includegraphics{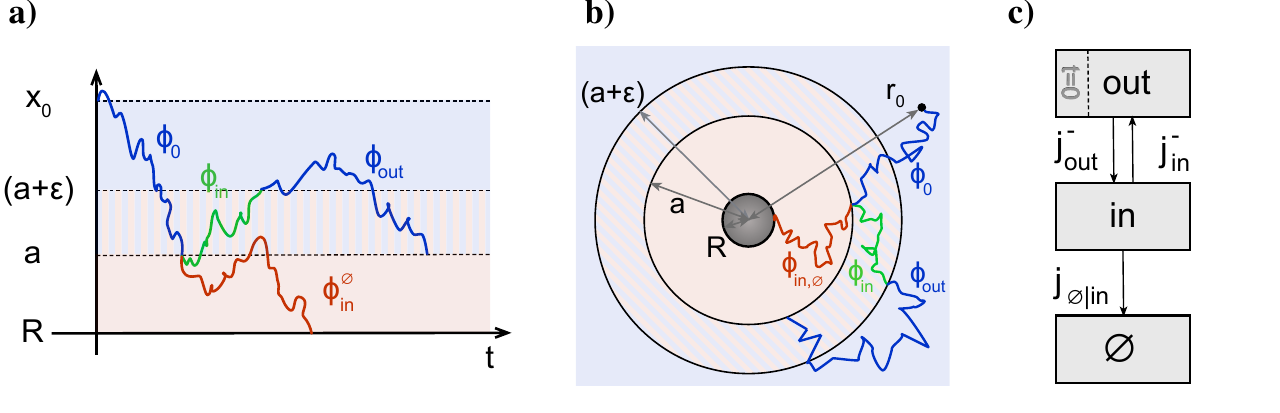}
	\caption{Geometries and exemplary partial trajectories of the first-passage problem with two domains as in \cref{fig:ChPn}, amended by an auxiliary boundary at $|\vec x|=a+\epsilon$. Panel~(c) shows the corresponding transition graph.
	Further details are given in the caption of \cref{fig:ChPn}.
	Note the hybrid character of the region $a < |\vec x| < a+\epsilon$: partial trajectories passing this region belong to either the $\In$ or the $\Out$ state, depending on the domain where they start.}
	\label{fig:ChP}
\end{figure}

As a regularization of these singularities, we require a minimum length $\epsilon > 0$ of the partial trajectories in the calculation of partial FPT distributions and we take the limit $\epsilon\to 0$ for the final result of the FPT distribution.
This can be accomplished by slightly shifting the boundary between $\In$ and $\Out$ for trajectories leaving the $\In$ state.
Technically, we introduce an auxiliary boundary at $|\vec x|= a+\epsilon$, with $a+\epsilon<b$ in the 3D case (\cref{fig:ChP}a,b).
This auxiliary boundary is transparent for partial trajectories in the $\Out$ state, i.e., they start in the outer domain and end at the boundary $|\vec x|=a$.
Partial trajectories starting at $|\vec x|=a$ are assigned to the $\In$ state, they either
end at $|\vec x| = a+\epsilon$ or are stopped at the absorbing boundary $|\vec x| = R$, whereas the boundary $a$ is invisible to the partial trajectories in the $\In$-state.
In the former case, the following partial trajectory belongs to the $\Out$ state, as do all
trajectories starting at $a + \epsilon$, and it ends at $|\vec x| = a$. Thus, the $\Out$ state is followed by an $\In$ state again and so forth. With the different states being assigned according to the different starting points, a partial trajectory cannot be successed by a partial trajectory in the same state,
as opposed to the situation in \cref{ssec:2dom}
(Reflections at the boundary $|\vec x| = b$ in the 3D case are included in the partial FPT density of the $\Out$ state and do not subdivide the partial trajectory further.)
The transport properties (e.g., diffusion coefficient) along a partial trajectory are those of the assigned state (or domain), which leads to a hybrid character of the region $a < |\vec x| < a + \epsilon$. Clearly, this interpretation is an approximation to the original diffusion problem, which is restored in the limit $\epsilon \to 0$.

Our definition of the occupation probabilities $\rho_\alpha(t)$ of the states $\alpha \in \{ \In, \Out, \emptyset \}$ remains unchanged. And as before, the probability fluxes $j_\alpha^{\pm}(t)$ denote gains ($+$) and losses ($-$) of the state $\alpha$;
the loss $j_{\emptyset|\In}$ from the $\In$ state to the absorbed state $\emptyset$ is not included in $j_\In^-(t)$.
The resulting transition graph of the amended two-domain GME is given in \cref{fig:ChP}.
In the terminology of \cref{ssec:GME}, the transmission probabilities at the $\In|\Out$ boundary are
$q_{\In|\Out} = q_{\Out|\In} = 1$, i.e., there are no transitions from a state to itself, $w_{\In|\In} = w_{\Out|\Out} = 0$, and the transitions between $\In$ and $\Out$ are thus strictly alternating. The loss fluxes do not split and flux continuity [\cref{eq:genfluxbal}] demands:
\begin{align}
j_{\In}^+(t) &=  j_{\Out}^-(t) \hspace{1cm} \text{and} \hspace{1cm}j_{\Out}^+(t) =  j_{\In}^-(t),
\label{fluxbal}
\end{align}
which replaces \cref{flbalinout1}.
For the local balance in each state, the same \cref{locbalout1,locbalin1,FPpdfnaive} as previously hold.
Also the \cref{flrenout1,flrenin1,flreninstop1} for the fluxes apply without modifications.
This set of equations constitutes the GME of the regularized two-domain model.

In the Laplace domain, the three expressions for the loss fluxes, \cref{flrenout1L,flrenin1L,flreninstop1L}, carry over as well since their derivation did not rely on \cref{flbalinout1}.
The linear system is completed by three equations for the occupation probabilities, which follow from \cref{fluxbal,locbalout1,locbalin1,FPpdfnaive}:
\begin{subequations}
\begin{align}
u \tilde\rho_\Out(u) - 1 &= \tilde j_\In^-(u) - j_\Out^-(u) \,, \label{rhooutLap}\\
u \tilde\rho_\In(u) &= \tilde j_\Out^-(u) - \tilde j_\In^-(u) - \tilde j_{\emptyset|\In}(u) \,, \label{rhoinLap}\\
u \tilde\rho_{\emptyset}(u) &= \tilde j_{\emptyset|\In}(u) \,. \label{rhoabsLap}
\end{align}
\end{subequations}
Solving the linear system in six variables, we find the desired FPT density
$\tilde p_{\text{FPT}}(u) = \tilde j_{\emptyset|\In}(u)$ in terms of the partial FPT densities:
\begin{equation}
\tilde p_{\text{FPT}}(u) =
  \frac{\tilde\phi_0(u) \, \tilde\phi_{\In}^{\emptyset}}{1 - \tilde\phi_\In(u) \, \tilde\phi_\Out(u)} \,,
\label{pFPT-L}
\end{equation}
which is a main result of this work. Note that the partial FPT densities [except $\tilde\phi_0(u)$] implicitly depend on the regularization parameter $\epsilon$.
The complete solution for all probabilities and fluxes is given in Eqs.~\eqref{eq:sol_extended} of the appendix.

\section{Diffusion in one dimension}\label{sec:1d}

The solution \eqref{pFPT-L} to the first-passage problem is completed by specifying the partial FPT densities for diffusion within each domain. Here and in the following section, we solve these FPT problems on homogeneous domains using standard techniques for the 1D and 3D geometries, respectively.

\subsection{FPT densities for partial trajectories}\label{sec:partFPT1d}

As mentioned above, the dwell time probability densities $\phi_\alpha(t)$ are themselves FPT densities, namely of the first passage from their entrance to the exit from the respective regions.
We will refer to these regions as ``outer'' and ``inner'' regions, respectively (see \cref{fig:ChP}).
These partial FPT densities can be obtained from the (backwards) Fokker--Planck equation for the probability density $\psi(\vec{x},t)$ of the particle position,
\begin{equation}
\frac{\partial}{\partial t}\psi(\vec{x},t)
=
D \frac{\partial^2}{\partial \vec{x}^2} \psi( \vec{x},t),\label{PDE}
\end{equation}
with an initial condition and the respective boundary conditions. 
More precisely, the setup where particles reaching some point in space for the first time are immediately removed from the system, corresponds to the setup in terms of position probability density with an absorbing boundary at that point. The flux into that point $\vec x_\xi$ will be the FPT density to visit that point for the first time,
\begin{equation}
\phi^\xi(\vec{x}_\xi,\vec{x}_\Omega,t;\vec{x_0}) = - D \frac{\partial}{\partial \vec{x}} \left.\psi(\vec{x},t)\right|_{\vec{x_\xi}},
\label{fluxdef}
\end{equation}
where the particles started at $\vec{x}=\vec{x_0}$ and $\vec x_\Omega$ is a point at the boundary.
% \textbf{how to write general flux, no lower \& upper BD yet?}

For each region we will solve the respective PDEs in Laplace domain where they take the form of ordinary differential equations linear in $\vec{x}$. The frequency $u$ will be kept as a variable, although it takes the role of a parameter during calculations.
In a first step, we will solve the pertinent initial-boundary-value-problems more generally for arbitrary starting points $x_0$ or $r_0$ and lower and upper boundaries at $x_\ell$, $x_u$ or $r_\ell$, $r_u$, respectively and customize them later. A more detailed exposition of the procedure than we are able to give here can be found in \cite{Redner:FirstPassage}.

Specializing to 1D spaces, we have the following equation governing the probability to find a particle at some $x$
\begin{equation}
 D \frac{\partial^2}{\partial x^2} \tilde\psi(x,u) - u \tilde\psi(x,u) = -\delta(x-x_0), \label{LapODE1d}
\end{equation}
where the right side represents the initial condition of all particles starting at $x_0$.
The fundamental system is
$
 \mleft\{\exp\mleft(\sqrt{u/D} \, x\mright),\exp\mleft(-\sqrt{u/D} \, x\mright)\mright\},
$
i.e. our solutions will be a linear combination of these two exponentials in $x$ and $\sqrt{u/D}$.

\subsubsection{Outer region}
In the outer region we have an absorbing lower boundary at some $x_\ell$ and a reflecting upper one at $x_u$:
 \begin{align}
\tilde\psi(x_\ell,u) &= 0 \,, \label{RB1do_1}\\
\frac{\partial}{\partial x}\left.\tilde\psi(x,u)\right|_{x=x_u} &= 0 \,, \label{RB1do_2}
\end{align}
for which
\begin{equation}
\tilde\psi(x,u) =\left\{
\begin{array}{cl}
\displaystyle
\frac{1}{\sqrt{uD}} \frac{\cosh\left(\sqrt{u/D}(x_u-x_0)\right)}{\cosh\left(\sqrt{u/D}(x_u-x_\ell)\right)}
\sinh\left(\sqrt{u/D}(x-x_\ell)\right), & x < x_0, \\
\displaystyle
\frac{1}{\sqrt{uD}} \frac{\sinh\left(\sqrt{u/D}(x_0-x_\ell)\right)}{\cosh\left(\sqrt{u/D}(x_u-x_\ell)\right)} 
\cosh\left(\sqrt{u/D}(x_u-x)\right), & x > x_0,
\end{array}
\right.
\end{equation}
solves \cref{LapODE1d}.
The fluxes onto the boundaries give us the FPT densities for a particle to reach the respective boundary.
Thus with
$
\tilde \phi^{\xi}(x_\ell,x_u,u;x_0) = \pm D \frac{\partial}{\partial x}\left.\tilde\psi(x,u)\right|_{x=x_\xi}
$
($+$ for the flux onto the lower and $-$ for the flux onto the upper boundary)
we have
\begin{align}
\tilde \phi^\ell_\Out(x_\ell,x_u,u;x_0) &= \frac{\cosh\left(\sqrt{u/D}(x_u-x_0)\right)}{\cosh\left(\sqrt{u/D}(x_u-x_\ell)\right)} \,, \\
\tilde \phi^u_\Out(x_\ell,x_u,u;x_0) &= 0.
\end{align}
The cumulative FPT probability to reach $x_\ell$ for $t\to\infty$ is obtained by putting $u\to 0$:
\begin{equation}
\int_0^\infty \phi^\ell_\Out(x_\ell,x_u,t;x_0) dt  = \lim_{u\to 0}\tilde \phi^\ell_\Out(x_\ell,x_u,u;x_0) = 1,
\end{equation}
as expected. To see what happens in an infinite system, we let
$x_u\to\infty$:
\begin{equation}
\lim_{x_u\to\infty}\tilde \phi^\ell_\Out(x_\ell,x_u,u;x_0) = \exp\left(-\sqrt{u/D}(x_0-x_\ell)\right),
\end{equation}
which is the known density for first passage to a point $x_\ell$ on an infinite domain.
An expansion in small $u$ yields a non-analytic expression which gives rise to long time tail. (In fact the exponential of a square root of $u$ gives the one sided L\'evy stable law $L_{\alpha}$ of parameter $\alpha=1/2$ in $t$, thus the tail is $\propto t^{-3/2}$.)

\subsubsection{Inner region}
Here we consider the boundary conditions
\begin{align}
\tilde\psi(x_\ell,u) &= 0 \,, \label{RB1di_1}\\
\tilde\psi(x_u,u) &= 0 \,, \label{RB1di_2}
\end{align}
for which the solution to \cref{LapODE1d} yields
\begin{equation}
\tilde\psi(x,u) =\left\{
\begin{array}{cl}
\displaystyle
\frac{1}{\sqrt{uD}} \frac{\sinh\left(\sqrt{u/D}(x_u-x_0)\right)}{\sinh\left(\sqrt{u/D}(x_u-x_\ell)\right)} \sinh\left(\sqrt{u/D}(x-x_\ell)\right), & x < x_0 , \\
\displaystyle
\frac{1}{\sqrt{uD}} \frac{\sinh\left(\sqrt{u/D}(x_0-x_\ell)\right)}{\sinh\left(\sqrt{u/D}(x_u-x_\ell)\right)} \sinh\left(\sqrt{u/D}(x_u-x)\right), & x > x_0 .
\end{array}
\right. 
\end{equation}
Again, we compute the fluxes onto the boundaries [cf.\ \cref{fluxdef}]:
\begin{align}
\tilde \phi^\ell_\In(x_\ell,x_u,u;x_0) &= \frac{\sinh\left(\sqrt{u/D}(x_u-x_0)\right)}{\sinh\left(\sqrt{u/D}(x_u-x_\ell)\right)} \,,\\
\tilde \phi^u_\In(x_\ell,x_u,u;x_0) &= \frac{\sinh\left(\sqrt{u/D}(x_0-x_\ell)\right)}{\sinh\left(\sqrt{u/D}(x_u-x_\ell)\right)} \,.
\end{align}
These fluxes are the FPT densities to the respective boundaries. The splitting probabilities, i.e. the cumulative probabilities to leave the system via 
the respective boundary, are obtained by sending $u\to0$,
which corresponds to $t\to \infty$ in time domain:
\begin{align}
P_{x_\ell}(x_0) &= \int_0^\infty \phi^\ell_\In(x_\ell,x_u,t;x_0) dt = \frac{x_u-x_0}{x_u-x_\ell} \,, \\
P_{x_u}(x_0) &= \int_0^\infty \phi^u_\In(x_\ell,x_u,t;x_0) dt = \frac{x_0-x_\ell}{x_u-x_\ell} \,.
\end{align}
Sending the outer boundary to infinity,
\begin{align}
\lim_{x_u\to\infty}\tilde \phi^\ell_\In(x_\ell,x_u,u;x_0)
&= \exp\left(-\sqrt{u/D}(x_0-x_\ell)\right) ,
\end{align}
we recover the known FPT density on an infinite domain.

\subsubsection{Scaling form and full solution for the FPT density}

Let now, closer in accordance with the original problem (\cref{fig:ChPn}a), start the partial trajectory at a small distance $\epsilon$ to its end point $a$, thus
$x_0 = a + \epsilon$, $x_\ell = a$ for $\tilde\phi_\Out$ and $x_0 = a$, $x_u = a + \epsilon$ for $\tilde\phi_\In$. The lower boundary for the $\phi^u_\In$ be $R$ and the upper boundary of the $\phi^\ell_\Out$ be $b$.
Furthermore, let us introduce the scaled Laplace variable
$s\coloneqq u \epsilon^2/D$. Then, we have
\begin{align}
\tilde\phi^u_\In(\epsilon,s) &= \frac{\sinh\left(\sqrt{s}(R-a)/\epsilon\right)}{\sinh\left(\sqrt{s}((R-a)/\epsilon - 1)\right)}\,, \label{phi_in_L1}\\
\tilde\phi^\ell_\Out(\epsilon,s) &= \frac{\cosh\left(\sqrt{s}((a-b)/\epsilon + 1)\right)}{\sinh\left(\sqrt{s}(a-b)/\epsilon\right)} \,.\label{phi_out_L1}
\end{align}
The FPT densities of the partial trajectories attain their scaling forms for $\epsilon\to 0$:
\begin{equation}
\lim_{\epsilon\to 0}\tilde\phi^{u}_{\In}(\epsilon,s)
  = \exp\mleft(-\sqrt{s}\mright) \label{sc-th-L},
\end{equation}
and the same for $\lim_{\epsilon\to 0}\tilde\phi^{\ell}_{\Out}(\epsilon,s)$, which permits an analytic inversion of the Laplace transform:
\begin{equation}
\frac{1}{2\sqrt{\pi} (t^*)^{3/2}} \exp\mleft(-\frac{1}{4t^*}\mright)
\label{sc-th}
\end{equation}
with $t^* :=  t D/\epsilon^2$ the scaled time variable conjugate to $s$.

\begin{figure}
  \includegraphics{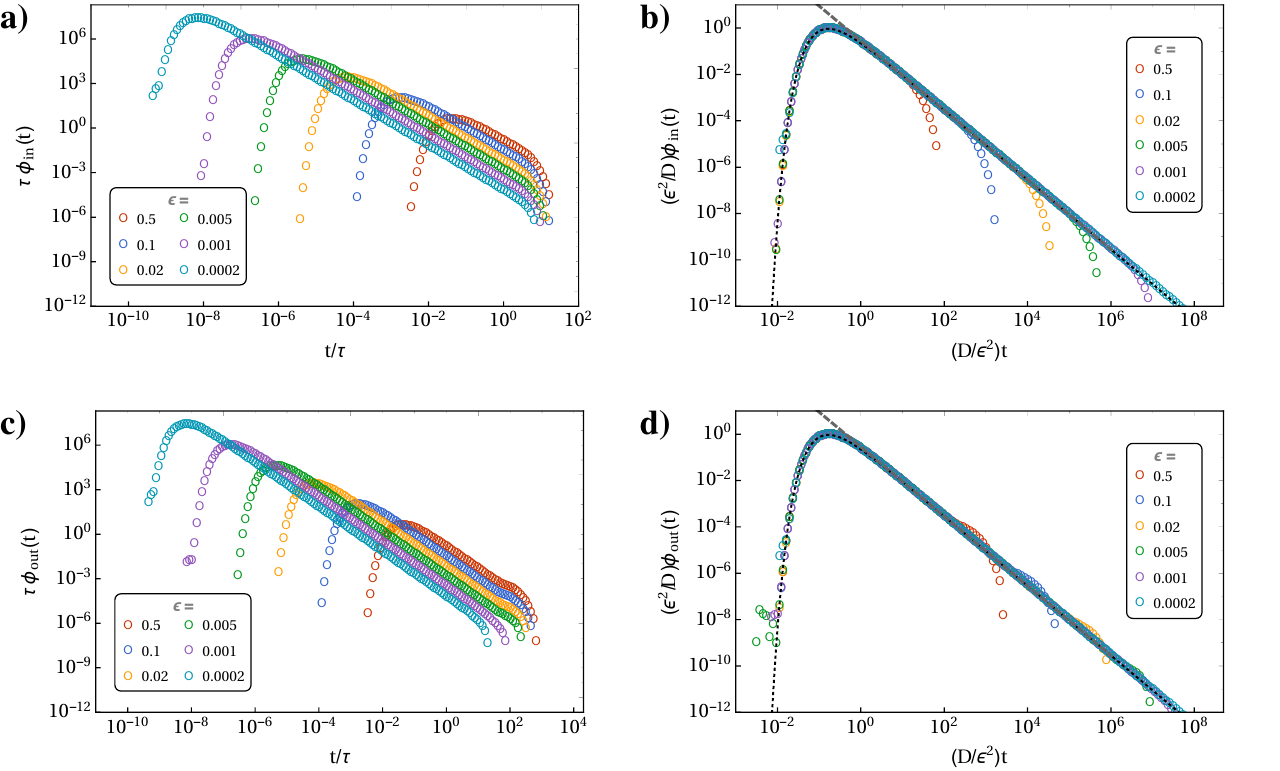}
	\caption{Numerical Laplace backtransforms (symbols) of the partial FPT densities $\phi_\In(t)$ in the inner domain [\cref{phi_in_L1}, panel a] and $\phi_\Out(t)$ in the outer domain [\cref{phi_out_L1} panel c] for the 1D problem.
  The panels b) and d) show the scaled partial FPT densities corresponding to the left panels.
		The parameters of the geometry are $x_0=10 R$, $a=5R$, and $b=20 R$, and $\tau=R^2/D$ is the unit of time.
		The black dotted line denotes the analytical solution in the limit $\epsilon\to 0$ [\cref{sc-th}].
		The gray dashed line indicates the asymptotic power law tail $t^{-3/2}$.}
	\label{fig:partFPT1d}.
\end{figure}

The partial FPT densities for the inner and outer domains are depicted in the left panels of \cref{fig:partFPT1d}.
The closer to the absorbing boundary or target a particle starts, the shorter the time of the peak position and the higher the peak value, indicating that the particles are more rapidly absorbed. Thus, moving the starting position closer to the target corresponds to a limiting procedure for the partial FPT densities, e.g., $\lim_{\epsilon \to 0} \phi_{\In}(\epsilon, t)  = \delta_+(t)$, converging to a singular peak at $t=0$:
if the particle starts at the position of the target, all probability is absorbed immediately within zero time.
% This $\delta$-peak is a pathological one, though, because still it exhibits a power law tail.
The construction of  an auxiliary $\epsilon$-shell around the boundary at $a$, see \ref{sec:GMEaux}, circumvents the difficulties arising from such singular FPT densities.
At intermediate times, we find the expected $t^{-3/2}$ power law decays for the particles exploring the outer domain yet without hitting the outer confinement. The FPT densities at large times decay rapidly due to the confinement.

For the inner domain, there is another reflecting boundary at $|\vec x| = a$. A sharp exponential cutoff sets in at times $t\approx a^2/D$ (\cref{fig:partFPT1d}a,b), which we attribute to partial trajectories that end as soon as they reach the outer boundary and do not contribute to the FPT density at longer times.
The outer domain has a reflecting boundary at $|\vec x| = b$, which also results in an exponential cutoff of the tail (\cref{fig:partFPT1d}c,d). However, the trajectories are continued and move on in their attempts to reach the target, which they will eventually do, but at a later time. This shifts and smears out the cutoff at large times. Moreover, due to the conservation of probability, the reflected trajectories, which in the case of an unbounded domain would have contributed into the power law tail, are responsible for the small shoulder of the FPT density at large times.
The right panels of the figures show the partial FPT densities scaled with respect to the distance of the starting point to the target $\epsilon$. Small values of $\epsilon$ are equivalent to a large domain size $b \gg a$, the particle feels the confinement at a later time and the differences between reflecting or absorbing outer boundary vanish.

\subsection{Full solution for the FPT density}

With the partial FPT densities calculated above, we are ready to write down the analytical expression for the Laplace transform of the full FPT density for a particle starting at $x_0$ and being absorbed at $R$ by substituting
\begin{align*}
\tilde\phi_0(u) &= \phi^\ell_\Out(x_\ell = a,x_u=b,u;x_0) \,, \\
\tilde\phi_{\In}^{\emptyset}(u) &= \phi^\ell_\In(x_\ell = R,x_u=a+\epsilon,u;x_0=a) \,, \\
\tilde\phi_\In(u) &=  \phi^u_\In(x_\ell = R,x_u=a+\epsilon,u;x_0=a) \,, \quad \text{and} \\
\tilde\phi_\Out(a,u) &= \phi^\ell_\Out(x_\ell = a,x_u=b,u;x_0=a+\epsilon)
\end{align*}
into \cref{pFPT-L}.

\begin{figure}
	\centering
	\includegraphics[width=\figwidth]{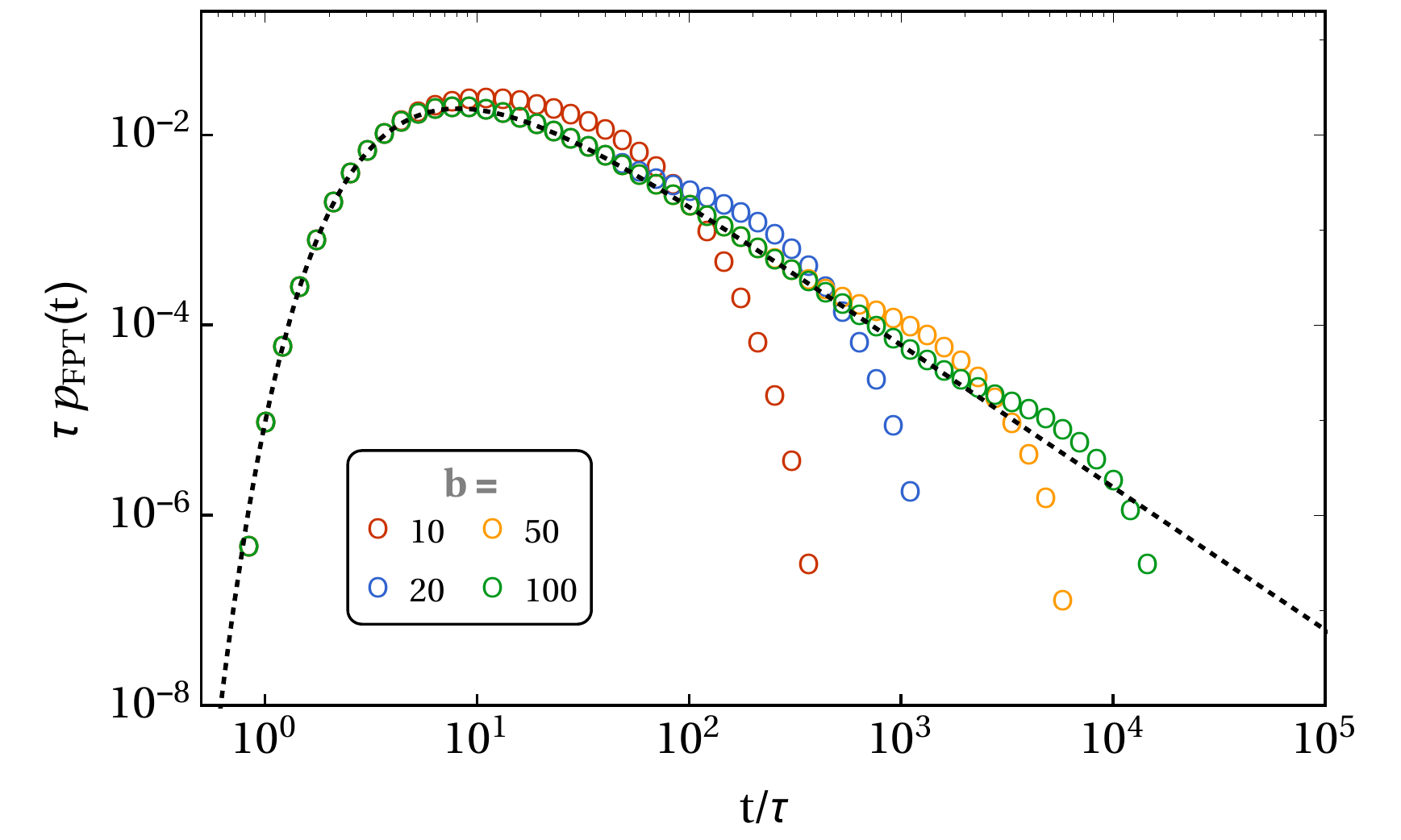}
	\caption{FPT density as obtained by numerical Laplace inversion of \cref{pFPT1d} for different domain sizes $b$ (in units of R), shown as symbols in different colors.
    Further parameters are $x_0/R=10$ and $a/R=5$, and $\tau = R^2/D$ is the unit of time.
		The black dotted line denotes the analytical solution in the limit $b\to \infty$ [\cref{pPFT1dth}]}
	\label{fig:totFPT1dpdf}
\end{figure}

As an ultimate test for the validity of the method
we compare our GME solution for the FPT density for the homogeneous system (i.e. particles behave the same way in inner and outer region) 
with the FPT density for particles starting at $r_0$ and ending at first encounter with $R$ (i.e. for non-dissected trajectories). 
We find perfect agreement between the two of them, 
\begin{align}
\tilde p_{\text{FPT}}(u) &=\tilde\phi^\ell_0(x_\ell=R,x_u=b,u;x_0) 
= \frac{\cosh\left(\sqrt{u/D}(b-x_0)\right)}{\cosh\left(\sqrt{u/D}(R-b)\right)}  \,.
\label{pFPT1d}
\end{align}
Observing that in this case, the $\epsilon$-dependence vanishes for the GME solution, taking the limit $\epsilon\to 0$ is not needed anymore. Also the dependence of $a$ vanishes as it should in the completely homogeneous case.
For an infinitely large outer domain, $b\to\infty$, we have:
\begin{align}
\tilde p^{\infty}_{\text{FPT}}(u) &= \exp\left(-\sqrt{u/D}\left(x_0 - R\right)\right) \,.
\label{FP1dinf}
\end{align}
In time domain, this limiting form corresponds to
\begin{align}
p^{\infty}_{\text{FPT}}(t) &= \frac{(x_0 - R)}{2 \sqrt{\pi D} t^{3/2}}
\exp\left(-\frac{(x_0 - R)^2}{4 D t}\right),
\label{pPFT1dth}
\end{align}
which is the expected result for first passage on an infinite domain.
As a last step, the inverse Laplace transform of expression \eqref{pFPT1d} is calculated numerically and depicted in \cref{fig:totFPT1dpdf}.
It is indeed qualitatively the same picture as already for the partial FPT densities with reflecting outer boundary: for small times, it fits the L\'evy--Smirnow law, \cref{pPFT1dth}, at large times, there is an exponential cutoff, and at FPTs smaller than the cutoff time, a small elevation relative to the $t^{-3/2}$ decay is visible, which collects the probability in the truncated tail.

\section{Diffusion in concentric spherical shells in 3D space}\label{sec:rad}

\subsection{FPT densities for the partial trajectory sections}\label{sec:partFPTrad}

Analogously to our calculations in \cref{sec:partFPT1d}, we now will calculate the partial FPT densities from the
fluxes onto the boundaries of the respective PDEs.
for the radial symmetric setup (\cref{fig:ChPn}b).
Thus in this case, \cref{PDE} becomes in Laplace domain:
\begin{equation}
 \frac{1}{r^2} \frac{\partial }{\partial r} r^2 D \frac{\partial }{\partial r} \tilde\psi(r,u) -  u \tilde\psi(r,u) =  - \frac{\delta(r-r_0)}{4\pi r_0} \,.
\end{equation}
By substituting $\tilde\eta(r,u) := \tilde\psi(r,u) / r$,
the equation for $\tilde\eta$ attains a similar form as the one in the 1D case:
 \begin{equation}
 D \frac{\partial^2}{\partial r^2} \tilde\eta(r) - u \tilde\eta(r) = -\frac{1}{4\pi r_0} \delta(r-r_0), \label{LapODErad}
 \end{equation}
with the initial condition specified on the right hand side (all trajectories start initially from the radius $r_0$).
The fundamental system is again
$
\mleft\{ \exp\bigl(\sqrt{u/D} \, r\bigr), \exp\bigl(-\sqrt{u/D} \, r\bigr) \mright\} \,.
$

\subsubsection{Outer region}
For the outer region, we have the transformed boundary conditions (absorbing at $r_\ell$, reflecting at $r_u$):
 \begin{align}
\tilde\eta(r_\ell,u) &= 0 \,,\label{RBro_1}\\
\frac{\partial}{\partial r}\left.\tilde\eta(r,u)\right|_{r=r_u} &= \frac{1}{r_u} \eta(r_u,u) \,, \label{RBro_2}
\end{align}
so that the solution to \cref{LapODErad} is given piecewiese for $r \lessgtr r_0$ by
\begin{multline}
\tilde\eta(r < r_0,u) =
\frac{1}{4\pi\sqrt{uD}r_0}
\frac{r_u\sqrt{\frac{u}{D}}\cosh\left(\sqrt{u/D}(r_0-r_u)\right) + \sinh\left(\sqrt{u/D}(r_0-r_u)\right)}
{r_u\sqrt{u/D}\cosh\left(\sqrt{u/D}(r_\ell-r_u)\right)
+ \sinh\left(\sqrt{u/D}(r_\ell-r_u)\right)} \\
\times \sinh\left(\sqrt{u/D}(r-r_\ell)\right)
\end{multline}
and
\begin{multline}
\tilde\eta(r > r_0,u) =
\frac{1}{4\pi\sqrt{uD}r_0}
\frac{\sinh\left(\sqrt{u/D}(r_0-r_\ell)\right)}
{r_u\sqrt{u/D}\cosh\left(\sqrt{u/D}(r_\ell-r_u)\right)
	+ \sinh\left(\sqrt{u/D}(r_\ell-r_u)\right)} \\
\times \left[r_u\sqrt{u/D}\cosh\left(\sqrt{u/D}(r-r_u)\right) + \sinh\left(\sqrt{u/D}(r-r_u)\right)\right] .
\end{multline}
With $\tilde\psi' = \tilde\eta^{\prime}/r - \tilde\eta/r^2$ and
$
\tilde \phi^{\xi}(r_\ell,r_u,u;r_0) = \pm 4D\pi r_\xi^2 \frac{\partial}{\partial r}\left.\tilde\psi(r,u)\right|_{r=r_\xi}
$
($+$ for the lower, $-$ for the upper boundary)
the fluxes onto the boundaries are thus
\begin{align}
\tilde \phi^\ell_\Out(r_\ell,r_u,u;r_0) &=
\frac{r_\ell}{r_0}  \frac{r_u\sqrt{u/D}\cosh\left(\sqrt{u/D}(r_0-r_u)\right) + \sinh\left(\sqrt{u/D}(r_0-r_u)\right)}
{r_u\sqrt{u/D}\cosh\left(\sqrt{u/D}(r_\ell-r_u)\right)
	+ \sinh\left(\sqrt{u/D}(r_\ell-r_u)\right)} \,, \\
\tilde \phi^u_\Out(r_\ell,r_u,u;r_0) &= 0 \,.
\end{align}
$\tilde \phi^\ell_\Out(r_\ell,r_u,u;r_0)$ is the density for first passage from $r_0$ onto $r_\ell$. % ($\tilde\phi_\Out(u)$ in our GME)

The integral over time is calculated as the limit $u\to 0$, which yields
%(by using $\sqrt{u/D}=v$ and l'H\^{o}spital's rule)
% have expression $0/0$!
%
\begin{equation}
\lim_{u\to0} \tilde\phi^\ell_\Out(r_\ell,r_u,u;r_0) = \lim_{v\to0} \left[
\frac{r_\ell}{r_0} \frac{r_u v\cosh\left(v(r_0-r_u)\right) + \sinh\left(v(r_0-r_u)\right) }{\cosh\left(v(r_\ell-r_u)\right) + \sinh\left(v(r_\ell-r_u)\right) }
\right] = \frac{r_\ell}{r_0} \frac{r_0}{r_\ell} = 1
\end{equation}
and confirms that with probability $1$ the particles ultimately reach the boundary $r_\ell$, as expected for a finite domain with the boundary at $r_\ell$ as the only exit.
We may consider the transition to an infinite domain by taking the limit of large $r_u$:
\begin{equation}
\lim_{r_u\to\infty} \tilde \phi^\ell_\Out(r_\ell,r_u,u;r_0)
  = \frac{r_\ell}{r_0} \exp\mleft(-\sqrt{u/D}(r_0-r_\ell)\mright) ,
\label{rotoinfrad}
\end{equation}
where again, as in the 1D case, the $\sqrt{u}$-term generates the long time tail.
Note that the limit $u\to 0$ in the infinite domain is less than unity, indicating the transience of diffusion in three dimensions.

\subsubsection{Inner region}
For the inner region, we have the boundary conditions:
\begin{align}
\tilde\eta(r_\ell,u) &= 0, \label{RBri_1}\\
\tilde\eta(r_u,u) &= 0, \label{RBri_2}
\end{align}
which results in
\begin{equation}
\tilde\eta(r,u) =\left\{
\begin{array}{cl}
\displaystyle
\frac{1}{4\pi\sqrt{uD}r_0}
\frac{\sinh\mleft(\sqrt{u/D}(r_u-r_0)\mright)}{\sinh\mleft(\sqrt{u/D}(r_u-r_\ell)\mright)}
\sinh\mleft(\sqrt{u/D}(r-r_\ell)\mright), & r < r_0 \\
\displaystyle
\frac{1}{4\pi\sqrt{uD}r_0}
\frac{\sinh\mleft(\sqrt{u/D}(r_0-r_\ell)\mright)}{\sinh\mleft(\sqrt{u/D}(r_u-r_\ell)\mright)}
\sinh\mleft(\sqrt{u/D}(r_u-r)\mright)
, & r > r_0.
\end{array}
\right.
\end{equation}
We find for the fluxes onto the boundaries:
\begin{align}
\tilde \phi^\ell_\In(r_\ell,r_u,u;r_0) &= \frac{r_\ell}{r_0}\frac{\sinh\mleft(\sqrt{u/D}(r_u-r_0)\mright)}{\sinh\mleft(\sqrt{u/D}(r_u-r_\ell)\mright)}
\\
\tilde \phi^u_\In(r_\ell,r_u,u;r_0) &= \frac{r_u}{r_0}\frac{\sinh\mleft(\sqrt{u/D}(r_0-r_\ell)\mright)}{\sinh\mleft(\sqrt{u/D}(r_u-r_\ell)\mright)}.
\end{align}
%
%Substituting $r_u \to (a+\epsilon)$ and $r_0 \to a$ 
%would provide us with $\tilde\phi_\In(u)$ and $\tilde\phi_{\In}^{\emptyset}(u)$, respectively.
\\
The time integrals, calculated as small-$u$ limits, provide the splitting probabilities:
\begin{align}
\lim_{u\to 0}\tilde \phi^\ell_\In(r_\ell,r_u,u;r_0) &= \frac{r_\ell}{r_0}\frac{r_u-r_0}{r_u-r_\ell} \,,\\
\lim_{u\to 0}\tilde \phi^u_\In(r_\ell,r_u,u;r_0) &= \frac{r_u}{r_0}\frac{r_0-r_\ell}{r_u-r_\ell} \,,
\end{align}
and indeed the sum of both is $1$ as it should.
Taking the limit $r_u\to\infty$ yields the same cumulative FP probability for an infinite system as in the case of a reflecting upper boundary [\cref{rotoinfrad}],
\begin{equation}
\lim_{r_u\to\infty}\tilde \phi^\ell_\In(r_\ell,r_u,u;r_0) = \frac{r_\ell}{r_0}
  \exp\mleft(-\sqrt{u/D} \, (r_0-r_\ell)\mright).
\end{equation}

\subsubsection{Scaling form}

Again, according to \cref{fig:ChPn}b, we let the partial trajectory start  at a small distance $\epsilon$ to its end point $a$, thus
$r_0 = a + \epsilon$, $r_\ell = a$ for $\tilde\phi_\Out$ and $r_0 = a$, $r_u = a + \epsilon$ for $\tilde\phi_\In$. The inner radius for the $\phi_\In$ be $R$ and the outer radius of the $\phi_\Out$ be $b$.
With the scaled Laplace variable $s\coloneqq u \epsilon^2/D$, we have
\begin{align}
\tilde \phi^u_\In(\epsilon,s) &= \frac{a+\epsilon}{a}\frac{\sinh\bigl(\sqrt{s}(R-a)/\epsilon\bigr)}{\sinh\bigl(\sqrt{s}[(R-a)/\epsilon - 1]\bigr)} \,,\label{phi_in_Lr}\\
\tilde \phi^\ell_\Out(\epsilon,s) &= \frac{a}{a+\epsilon}
\frac{b\sqrt{s}/\epsilon \cosh\bigl(\sqrt{s}[(a-b)/\epsilon + 1]\bigr) + \sinh\bigl(\sqrt{s}[(a-b)/\epsilon + 1]\bigr)}
{b\sqrt{s}/\epsilon \cosh\bigl(\sqrt{s}(a-b)/\epsilon\bigr) + \sinh\bigl(\sqrt{s}(a-b)/\epsilon\bigr)} \,.
\label{phi_out_Lr}
\end{align}
The limits as $\epsilon\to 0$ are equal to the scaling forms of the 1D case [\cref{sc-th-L,sc-th}].

\begin{figure}
  \includegraphics{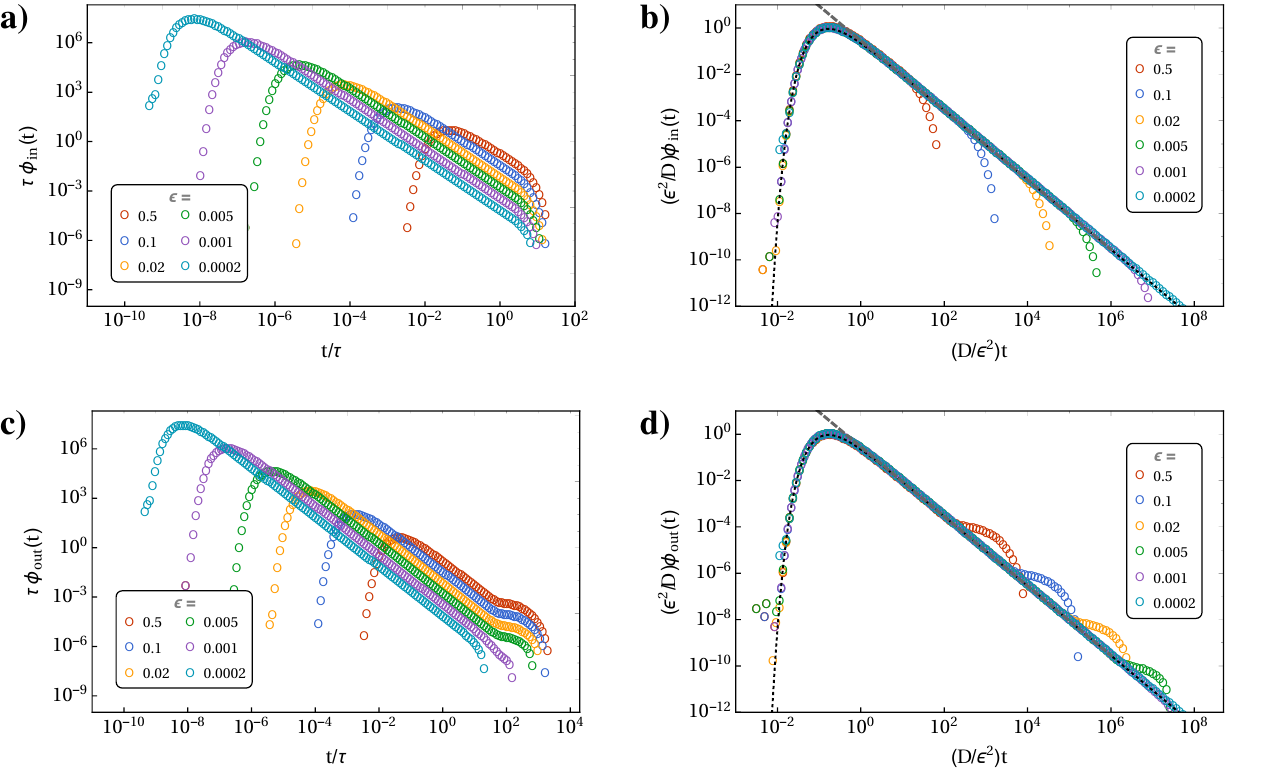}
	\caption{Numerical Laplace backtransforms (symbols) of the partial FPT densities $\phi_\In(t)$ in the inner shell [\cref{phi_in_Lr}, panel~a] and $\phi_\Out(t)$ in the outer shell [\cref{phi_out_Lr}, panel~c] for the radial problem in 3D.
		The panel b) and d) show the scaled partial FPT densities corresponding to the left panels.
		The parameters of the geometry are $r_0=10R$, $a=5R$, and $b=20R$, and $\tau=R^2/D$ is the unit of time.
		The black dotted line denotes the analytical solution in the limit $\epsilon\to 0$ [\cref{sc-th}].
		The gray dashed line indicates the asymptotic power law tail $t^{-3/2}$.}
  \label{fig:partFPTrad}
\end{figure}

\Cref{fig:partFPTrad} shows the partial FPT densities for the inner and outer domains (panels a,c).
Similarly to the 1D case we find again the peak at small times, which shifts to the left and gets the higher (and narrower) the closer to the target the particle starts.
Intermediate times are again governed by a $t^{-3/2}$ power law decay. At large times, the cutoff sets in, again
relatively sharply at $t \approx a^2/D$ for the inner domain with its absorbing outer boundary at $|\vec x|=a$, and more blurred out near $t\approx b^2/D$ for the outer domain with its reflecting boundary at $|\vec x|=b$. In the 3D case, the small shoulder at large times is more pronounced as compared to the 1D case. This is a well-known effect and is due to the compact exploration of space in low dimensions, where the first passage is more dominated by the direct trajectories \cite{Benichou:2014}. A smaller large-time shoulder indicates that the geometry plays a minor role in low dimensions than in 3D or higher.
Data collapse of the partial FPT densities is demonstrated again by rescaling with respect to the distance $\epsilon$ of the starting point to the target (\cref{fig:partFPTrad}b,d).

\subsection{Full solution for the FPT density}

Again we use the partial FPT densities to obtain the Laplace transform of the full FPT densities for a particle starting at $r_0$ and being absorbed at $R$
by substituting
\begin{subequations}
\label{eq:phi-results}
\begin{align}
\tilde\phi_0(u) &= \phi^\ell_\Out(r_\ell = a,r_u=b,u;r_0) \,, \\
\tilde\phi_{\In}^{\emptyset}(u) &= \phi^\ell_\In(r_\ell = R,r_u=a+\epsilon,u;r_0=a) \,,\\
\tilde\phi_\In(u) &=  \phi^u_\In(r_\ell = R,r_u=a+\epsilon,u;r_0=a) \,, \quad \text{and}\\
\tilde\phi_\Out(a,u) &= \phi^\ell_\Out(r_\ell = a,r_u=b,u;r_0=a+\epsilon)
\end{align}
\end{subequations}
into \cref{pFPT-L}.

%e FPT density for particles starting at $r_0$ and ending at first encounter with $R$ (i.e. for non-dissected trajectories)

The coincidence of our solution for the FPT density for the homogeneous system obtained via the GME approach with the time density for first passage to $R$ starting from $r_0$, again underlines its vanishing dependence on $\epsilon$ and $a$,
\begin{align}
\tilde p_{\text{FPT}}(u) &= \tilde\phi^\ell_0(r_\ell=R,r_u=b,u;r_0) \nonumber\\
&=\frac{R}{r_0}  \frac{b\sqrt{u/D}\cosh\left(\sqrt{u/D}(r_0-b)\right) + \sinh\left(\sqrt{u/D}(r_0-b)\right)}{b\sqrt{u/D}\cosh\left(\sqrt{u/D}(R-b)\right)},
\label{pFPTrad}
\end{align}
as expected.
For an infinite outer domain, $b\to\infty$, we have:
\begin{align}
\tilde p^{\infty}_{\text{FPT}}(u) &=\tilde\phi^{\infty}_0(r_\ell=R,u;r_0)
=\frac{R}{r_0}\exp\left(-\sqrt{u/D}\left(r_0 - R\right)\right),
\label{FPradinf}
\end{align}
which is translated to the time domain as
\begin{align}
p^{\infty}_{\text{FPT}}(t) &=\frac{R}{r_0}\frac{(r_0 - R)
}{2 \sqrt{\pi D} t^{3/2}}\exp\left(-\frac{(r_0 - R)^2}{4 D t}\right).
\label{pPFTradth}
\end{align}

\begin{figure}
	\centering
	\includegraphics[width=\figwidth]{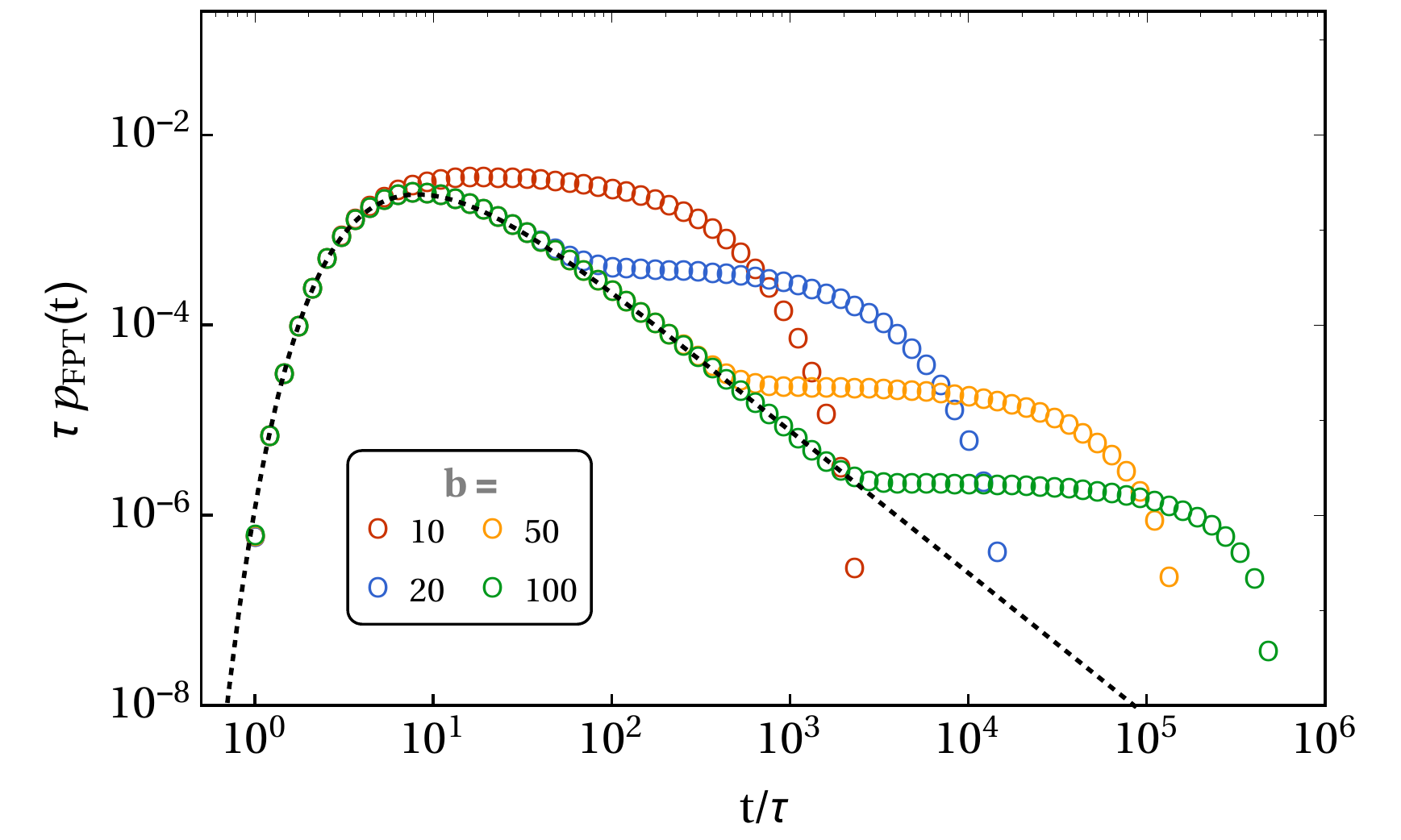}
	\caption{FPT densities as obtained by numerical Laplace inversion of \cref{pFPTrad} for different domain sizes $b$ (in units of $R$), shown as symbols of different color.
	Further parameters are $r_0/R=10$, $a/R=5$, and $\tau=R^2/D$ is the unit of time.
	The black dotted line denotes the analytical solution in the limit $b\to \infty$, \cref{pPFTradth}.}
	\label{fig:totFPTradpdf}
\end{figure}

The numerical Laplace inversion of \cref{pFPTrad} yields the final FTP densities $p_\FPT(t)$, shown in \cref{fig:totFPTradpdf} and deviating qualitatively from the 1D case (\cref{fig:totFPT1dpdf}): the plateau region is much more pronounced in the 3D case. Notice that the limiting FPT density for infinite domains [\cref{pPFTradth}] does not normalize due to its prefactor, while the FPT densities in confined domains do. The plateau in the FPT profile for reflecting boundaries does not only compensate for those particles that would have contributed to the power law tail in the case of an unbounded domain, but in addition it compensates also for the particles that would have got lost forever due to the transient character of diffusion in dimensions higher than two.

\section{Application: domains with different diffusivity}
\label{sec:application}

As an application to diffusion in a non-uniform, piecewise homogeneous medium made of two concentric spherical shells, we modify the above calculations for the 3D case and assign distinct diffusion constants, $D_\In$ and $D_\Out$, to the inner and outer domains, respectively; such a geometry was studied in Ref.~\cite{Godec:SR2016}.
The modified diffusion constants enter merely the partial FPT densities, so that the same derivations as before go through.
No further constraints or boundary conditions are needed in the present framework: the transitions between the domains are fully determined by the condition of flux continuity at the domain boundaries, \cref{fluxbal}.
Thus, $\tilde p_\FPT(u)$ obeys \cref{pFPT-L}, but the explicit result will be different from \cref{pFPTrad}.

In this setup, it is also interesting to consider the case where the particle starts in the inner region, $R<r_0<a$.
As pointed out in Ref.~\cite{Godec:SR2016}, the FPT density in this case is governed by a third timescale which manifests in an additional intermediate regime in the FPT density.
When the particle starts on the inside, the governing equations change slightly due to the different initial conditions, $\rho_\In(0)=1$, $\rho_\Out(0)=0$. \Cref{fluxbal,locbalout1,locbalin1,FPpdfnaive} remain the same, but for the recursive equations for the loss fluxes we have:
\begin{subequations}
	\begin{align}
	j_\Out^- (t) &=  \int_0^t \phi_\Out(t-t^\prime) j^+_\Out(t^\prime) dt^\prime \,, \label{flrenouti}\\
	j_\In^-(t) &= \phi_{0}^{\In}(t)\rho_\In(0) + \int_0^t \phi_\In(t-t^\prime) j^+_\In(t^\prime) dt^\prime  \,,\label{flrenini}\\
	j_{\emptyset|\In}(t) &= \phi_{0}^{\emptyset}(t)\rho_\In(0) + \int_0^t \phi_{\In}^{\emptyset}(t-t^\prime) j^+_\In(t^\prime) dt^\prime  \,,\label{flreninstopi}
	\end{align}
\end{subequations}
where $\phi_{0}^{\In}$ and $ \phi_{0}^{\emptyset}(t)$ are the dwell time probability density for a particle on an inner partial trajectory starting from $r_0$ and ending at $(a+\epsilon)$ or $R$, respectively.
In Laplace domain, the linear system of equations consists of
\begin{subequations}
	\begin{align}
	\tilde j_\In^-
	&= \frac{\tilde\phi_\In(u)}{1-\tilde\phi_\In(u)}
	\left[u \tilde\rho_\In(u) + \tilde j_{\emptyset|\In}(u)\right] ,
	\label{jiniLap}\\
	\tilde j_\Out^-
	&= \frac{\tilde\phi_0(u) - \tilde\phi_\Out(u) }{1-\tilde\phi_\Out(u)}
	+\frac{\tilde\phi_\Out(u) }{1-\tilde\phi_\Out(u)}
	u \tilde\rho_\Out(u) \,,
	\label{joutiLap}\\
	\tilde j_{\emptyset|\In}
	&= \frac{\tilde\phi_{\In}^{\emptyset}(u)}{1-\tilde\phi_{\In}^{\emptyset}(u)}
	\left[u \tilde\rho_\In(u) + \tilde j_\In^-(u)\right]
	\label{jabsiLap}
	\end{align}
\end{subequations}
for the loss fluxes, and
\begin{subequations}
	\begin{align}
	u \tilde\rho_\Out(u) &= \tilde j_\In^-(u) - j_\Out^-(u) \,, \label{rhooutiLap}\\
	u \tilde\rho_\In(u) - 1 &= \tilde j_\Out^-(u) - \tilde j_\In^-(u) - \tilde j_{\emptyset|\In}(u) \,, \label{rhoiniLap}\\
	u \tilde\rho_{\emptyset}(u) &= \tilde j_{\emptyset|\In}(u) \label{rhoabsiLap}
	\end{align}
\end{subequations}
for the probabilities that a domain is occupied (i.e., a particle being inside of radius $a$, outside or stopped),
from which we extract the FPT density in Laplace domain (see appendix for the full solution):
\begin{equation}
\tilde p_\FPT(u) = \tilde j_{\emptyset|\In} = \tilde\phi_{0}^{\emptyset}(u) +
\frac{\tilde\phi_{0}^{\In}(u)\tilde\phi_{\In}^{\emptyset}(u)\tilde\phi_\Out(u)}
{1 -  \tilde\phi_\In(u) \tilde\phi_\Out(u)}. \label{pFPT-L_in}
\end{equation}
For both cases of the particle starting from outside or within the radius $a$, the $\phi$ functions as calculated in \cref{sec:rad} were inserted into \cref{pFPT-L} and \cref{pFPT-L_in} with the respective diffusion constants. Finally, the resulting expressions were Laplace-inverted numerically.

\begin{figure*}
	\includegraphics{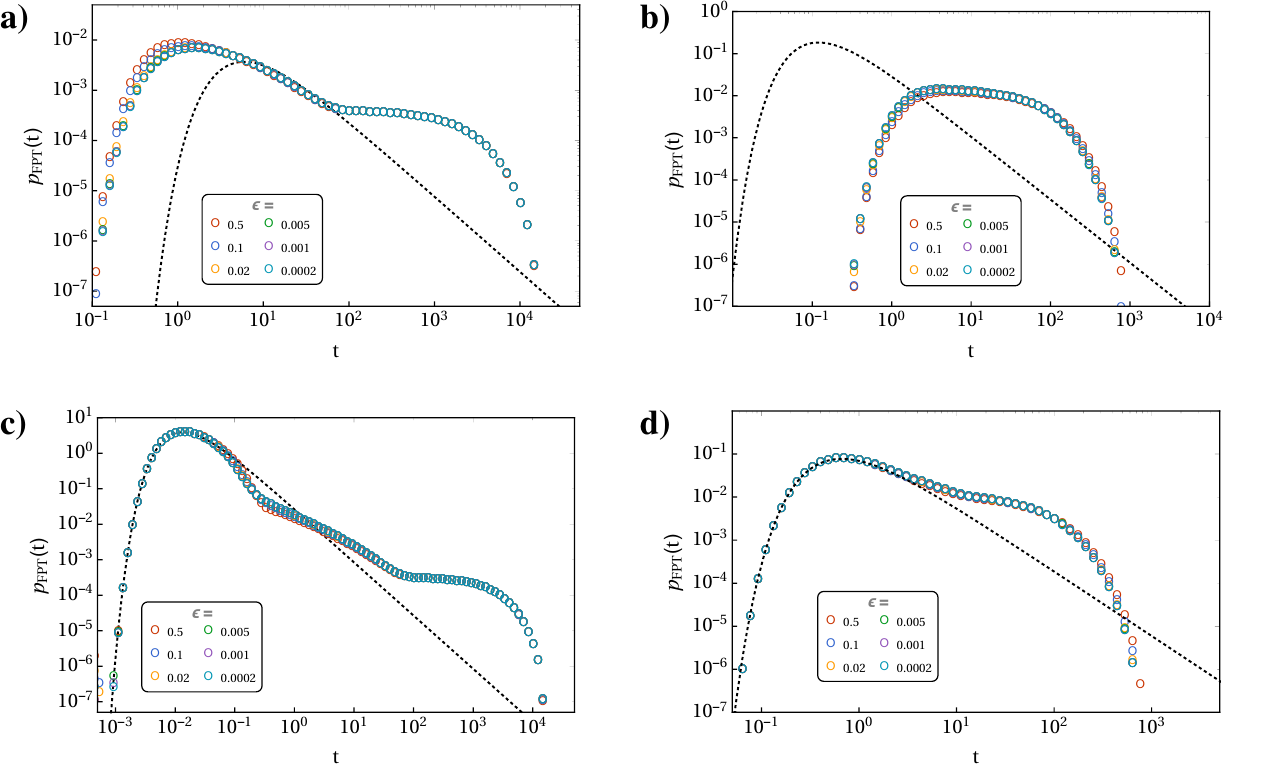}%
	\caption{FPT densities for the radial setup with distinct diffusion constants in the inner and outer regions.
	Diffusion is fast, relative to the homogeneous setup, either in the inner region $D_\In = 50 D$ and $D_\Out =D$ (panels a,c) or in the outer region $D_\In =D$ and $D_\Out = 50 D$ (panels b,d).
	Particles start either from the outer domain (parameters $r_0/R=10$ and $a/R=5$, panels a,b) or from the inner domain ($r_0/R=6$ and $a/R=8$, panels c,d); in all cases, the outer radius is $b/R=20$.
	The data were obtained from \cref{pFPT-L} with Eqs.~\eqref{eq:phi-results} by numerical inversion of the Laplace transform.
	For comparison, dotted lines indicate the analytical solution of the homogeneous problem in the limit $b\to \infty$ [\cref{pPFTradth}] setting $D=D_\Out$ for the upper panels (a,b) and $D=D_\In$ for the lower panels (c,d), respectively.
	}
	\label{fig:inhradFPTpdf}
\end{figure*}

In contrast to the homogeneous case, the $\epsilon$- and $a$-dependence of the FPT prevails,
but the limit $\epsilon\to 0$ always exists. As seen from \cref{fig:inhradFPTpdf}, already for moderate $\epsilon$ there is hardly an influence on the total FPT density.
Obviously the  overall time spent in the region  $(a,a+\epsilon)$ is negligible compared to the time spent in the inner and outer region so that the accumulated relative error made by constructing the auxiliary $\epsilon$-shell around $a$ remains small.
As a comparison and for a qualitative discussion, we have also included the respective FPT density for an infinite domain in the FPT plots: \cref{pPFTradth} with $D=D_\Out$ for the particle starting outside (\cref{fig:inhradFPTpdf}a,b) and $D=D_\In$ for the particle starting inside of the radius~$a$ (\cref{fig:inhradFPTpdf}c,d).

In the case of the particles starting outside (\cref{fig:inhradFPTpdf}a,b), we see either an enhancement or a delay in the left peak indicating the particles that proceed directly to the target $R$, depending on whether the diffusion in the inner region is fast (panel~a) or slow (panel~b). Moreover, we have a broader tail in the case where particles are slower in the outer region (panel~a). It takes a longer time for the particle to traverse the whole domain before the exponential cutoff due to the finiteness of the accessible space comes into effect.
In the case of the particles starting from within the radius $a$, we infer from \cref{pFPT-L_in} that the distribution is a superposition of the FPT distribution of the direct trajectories $\tilde\phi_{0}^{\emptyset}(u)$ and another term. For small times, $\tilde\phi_{0}^{\emptyset}(u)$ is well approximated by the FPT density on an infinite domain. With respect to the broadening of the FPT density for small $D_\Out$ we have the same effect as for particles starting outside, but in addition, an intermediate regime emerges for those particles that initially leave the inner region, but then proceed to the target without much exploring the outer region. A detailed analysis of the various regimes based on a completely different method can be found in Ref.~\cite{Godec:SR2016}, albeit the boundary conditions used there are slightly different from this work.
The close similarity of their FPT densities with our findings underlines that the
present approach is suitable to yield FPT densities for radially symmetric domains with distinct diffusion constants and a central target.

% \Cref{fig:inhradFPTpdf}c,d

\section{Summary and conclusions}

We have introduced a novel framework to stochastic first-passage problems in non-uniform, piecewise homogeneous environments, which is suitable to yield FPT distributions, but also splitting probabilities and total sojourn times in a domain.
The central requirement is the Markov property of the underlying transport process, which allowed us to cast the FPT problem into a renewal problem.
To this end, we coarse-grained the diffusion trajectories by dissecting them at the crossing points between adjacent domains, for example, between regions of different diffusivities; each partial trajectory is interpreted as a state, labeled by $\alpha$ (\cref{fig:ChPn}).
The dwell times on each domain are determined by the dynamics in the domain, e.g., by the time it takes a molecule to leave the domain. The dwell time distributions $\phi_{\alpha|\beta}^\gamma(t)$ parameterize the coarse-grained model and summarize the geometry, dimensionality, diffusion properties, etc. of the respective domain.
They may be obtained from solving first-passage problems on the (homogeneous) domains, or from simulations or even experiments.
In general, the dwell times are not exponentially distributed with the consequences that the coarse-grained stochastic process is not a Markov jump process and that the evolution of the occupation probabilities $\rho_\alpha(t)$ does not obey a classical master equation.
The latter is replaced by a generalized master equation (GME), which is a set of linear integro-differential equations for the probabilities $\rho_\alpha(t)$ and their fluxes, thereby accounting for memory effects.

A subtle difficulty of the approach is the singular character of first-return times to the same domain boundary in a continuous space. We showed that this issue can be solved by assigning a finite width $\epsilon$ to the domain boundaries where needed (\cref{fig:ChP}), which regularizes the partial FPT densities; the limit $\epsilon \to 0$ is then taken in the final results.

As a test case, we exemplified the GME approach for two-domain models in 1D and 3D with domains of equal diffusivities and showed that the known FPT distributions $p_\FPT(t)$ are recovered; here, the dependence on $\epsilon$ dropped out.
An analytical solution of the GME is readily obtained in frequency domain by a Laplace transform [\cref{pFPT-L}]
with explicit results for the overall FPT density [\cref{pFPT1d,pFPTrad}].
We have complemented these results by a numerical backtransform to the temporal domain using an algorithm \cite{Straube:CP2020} that has proven robust for broad FPT densities extending over many decades (\cref{fig:totFPT1dpdf,fig:totFPTradpdf}).
Based on these results, it was straightforward to address a 3D target search problem with two domains of distinct diffusivities. The obtained FPT densities (\cref{fig:inhradFPTpdf}) exhibit a complex structure, characterized by three time scales, and resemble the findings of Ref.~\cite{Godec:SR2016}.

Having the validity of the GME approach corroborated, we are now in a position to investigate first-passage problems in other, more complex scenarios.
The modular approach of the method makes it particularly simple to extend the discussed two-domain models to heterogeneous spaces formed by a larger number of domains, including layered hetero-structures, or to assemble models for the interplay of different modes of transport, giving scope for a variety of problems of heterogeneous diffusion in space and also in time.
On the other hand, it enables us to trace back the origin of certain features that emerge, e.g., in the FPT densities, since the results in frequency domain are always algebraic compositions of the FPT densities of the partial trajectories, reflecting the properties of the respective spatial domain (or transport mode).
There are various applications, where an ensemble of molecules diffuses and the first passage of \emph{some} molecule is relevant.
Such a scenario can be deduced from the single-particle solutions given here by adopting ideas of ref.~\cite{Grebenkov:NJP2020}, which appears as a feasible program as long as the particles are independent, i.e., they do not interact while they diffuse.

Manifestations of domain-specific transport behavior that can easily be accounted for in our framework on the level of the partial FPT densities are, e.g., degradation (or even reaction) of molecules,
different dimensionalities of the space of motion,
and different types of trapping subdiffusion.
One could also implement crowding effects by using models of anomalous diffusion in certain domains, albeit this often introduces memory on the trajectory level that would be lost when molecules leave a domain.
However, such a loss of memory may be justified if the transition between domains is associated with barrier crossing, e.g., due to a cellular membrane.
Barrier crossing and directed channeling can be modeled in our framework by altering the flux balance equations and by, e.g.,
modifying the splitting of probability at the domain interfaces or by adding a bias in the transitions.
Leaving these more complex situations for future work,
we conclude that the presented GME-based framework is a potentially powerful toolbox to address first-passage related questions of heterogeneous diffusion processes for a variety of transport modes.

\begin{acknowledgments}
We acknowledge the support of Deutsche Forschungsgemeinschaft through 
the Collaborative Research Center SFB~1114
``Scaling Cascades in Complex Systems'' (project ID: 235221301, subproject C03)
and under Germany's Excellence Strategy -- MATH+ : The Berlin Mathematics
Research Center (EXC-2046/1) -- project ID: 390685689, subproject EF4-4.
The data that support the findings of this study are available upon reasonable
request from the authors.
\end{acknowledgments}

\appendix

\section{Derivation of the simplified renewal equation \eqref{eq:genlossflren}}
\label{sec:simplifiedGME}

The starting point of the derivation of \cref{eq:genlossflren} is the general renewal equation for the probability flux $j_{\alpha|\beta}$ from domain $\beta$ to domain $\alpha$, which is repeated here for convenience:
\begin{equation}
 j_{\alpha|\beta} = q_{\alpha|\beta}\sum_{\gamma\neq\beta} \phi_{\beta|\gamma}^\alpha *
  \mleft[j_{\beta|\gamma} + (q_{\gamma|\beta}^{-1} - 1) j_{\gamma|\beta}\mright]
  + q_{\alpha|\beta} \,\phi_{\beta|0}^\alpha \,\rho_\beta^{(0)} \,; \quad \alpha \neq \beta.
  \tag{\ref{eq:genflren}}
\end{equation}
Suppose a sufficiently symmetric domain $\beta$ such that the partial FPT densities can be replaced
by $\phi_\beta^{(d)} = \phi_{\beta|\gamma}^\alpha$ for transitions between distinct boundaries and
$\phi_\beta^{(s)} = \phi_{\beta|\alpha}^\alpha$ for transitions from the boundary $\beta|\alpha$ to itself.
Further, all transmission probabilities are equal, $q_{\alpha|\beta} = q_\beta$, which implies
\begin{equation}
  j_{\beta|\beta} = (1- q_\beta) j_\beta^- \,.
  \label{eq:aux1}
\end{equation}

Under these assumptions, the summation of \eqref{eq:genflren} over the domains $\alpha$ that are adjacent to $\beta$ can be carried out.
For the left hand side, one finds
\begin{equation}
 \sum_{\alpha\neq\beta} j_{\alpha|\beta} = j_\beta^- - j_{\beta|\beta} = q_\beta j_\beta^- \,.
\end{equation}
For the self-term in the $\gamma$-sum on the r.h.s. ($\gamma = \alpha$), we calculate
\begin{equation}
\begin{split}
  \mathrlap{\phi_\beta^{(s)} * \sum_{\alpha\neq \beta} [q_\beta j_{\beta|\alpha} + (1 - q_\beta) j_{\alpha|\beta}]}
  \hspace{5em} \\
  &= \phi_\beta^{(s)} * [q_\beta (j_\beta^+ - j_{\beta|\beta}) + (1 - q_\beta) (j_\beta^- - j_{\beta|\beta})] \\
  &= q_\beta \phi_\beta^{(s)} * j_\beta^+ \,,
\end{split}
\end{equation}
using \cref{eq:aux1} to cancel $j_\beta^-$ and $j_{\beta|\beta}$.
For the distinct part, we consider first
\begin{equation}
\begin{split}
  \mathrlap{\sum_{\gamma\neq \alpha,\beta} [q_\beta j_{\beta|\gamma} + (1 - q_\beta) j_{\gamma|\beta}]}
  \hspace{5em} \\
  &= q_\beta (j_\beta^+ - j_{\beta|\alpha} - j_{\beta|\beta}) + (1 - q_\beta) (j_\beta^- - j_{\alpha|\beta} - j_{\beta|\beta}) \\
  &= q_\beta j_\beta^+ - q_\beta j_{\beta|\alpha} - (1 - q_\beta) j_{\alpha|\beta},
\end{split}
\end{equation}
making use of \cref{eq:aux1} again, and conclude that
\begin{equation}
\begin{split}
  \mathrlap{\phi_\beta^{(d)} * \sum_{\alpha\neq \beta} \sum_{\gamma\neq \alpha,\beta} [q_\beta j_{\beta|\gamma}
    + (1 - q_\beta) j_{\gamma|\beta}]}
  \hspace{5em} \\
  &= z_\beta q_\beta \phi_\beta^{(d)} * j_\beta^+ - \phi_\beta^{(d)} * \sum_{\alpha\neq \beta} [q_\beta j_{\beta|\alpha} + (1 - q_\beta) j_{\alpha|\beta}] \\
  &= z_\beta q_\beta \phi_\beta^{(d)} * j_\beta^+ - \phi_\beta^{(d)} * [q_\beta (j_\beta^+ - j_{\beta|\beta}) + (1 - q_\beta) (j_\beta^- - j_{\beta|\beta})] \\
  &= (z_\beta - 1) q_\beta \phi_\beta^{(d)} * j_\beta^+ \,,
\end{split}
\end{equation}
where $z_\beta$ is the number of domains next to~$\beta$.
For the initial term in \eqref{eq:genflren}, the summation is only over the FPT densities and we introduce the overall initial FPT density $\phi_\beta^{(0)} := \sum_{\alpha\neq\beta} \phi_{\beta|0}^\alpha$.
Collecting terms, the factor $q_\beta$ cancels and one recovers \cref{eq:genlossflren} as claimed:
\begin{equation}
 j_\beta^- = \phi_\beta * j_\beta^+ + \phi_\beta^{(0)} \rho_\beta^{(0)}
\end{equation}
with the overall FPT density
$\phi_\beta := \phi_\beta^{(s)} + (z_\beta - 1) \phi_\beta^{(d)}$
for reaching some point on the domain boundary of~$\beta$.

\section{Full solutions of the GME in Laplace domain}

The GME for the two-domain model was given in the Laplace domain by \cref{flrenout1L,flreninstop1L,flrenin1L,urhooutnL,urhoinnL,urhostopnL}. The solution for the three probabilities and the three fluxes read as follows upon abbreviating the common expression
$\tilde\chi_1(u) = 1 - \tilde\phi_\In(u) [1 - \tilde\phi_{\In}^{\emptyset}(u) / 2]$:
\begin{subequations}
\label{eq:sol_naive}
\begin{align}
\tilde\rho_\In(u) &= \frac{\tilde\phi_0(u)}{u}
  \frac{\tilde\chi_1(u) - \tilde\phi_{\In}^{\emptyset}(u)}
  {1 - \tilde\phi_\Out(u) + \tilde\chi_1(u)} \,, \\
\tilde\rho_\Out(u) &= \frac{1}{u} - \frac{\tilde\phi_0(u)}{u} \frac
 {\tilde\chi_1(u)}
  {1 - \tilde\phi_\Out(u) + \tilde\chi_1(u)} \,, \\
\tilde\rho_{\emptyset}(u) &= \frac{\tilde\phi_0(u)}{u} \frac
 {\tilde\phi_{\In}^{\emptyset}(u)}
  {1 - \tilde\phi_\Out(u) + \tilde\chi_1(u)}, \\
\tilde j_\In^-(u) &= \tilde\phi_0(u) \frac
 {\tilde\chi_1(u) - 1}
 {1 - \tilde\phi_\Out(u) + \tilde\chi_1(u)} \,, \\
\tilde j_\Out^-(u) &= \tilde\phi_0(u) \frac
 {\tilde\chi_1(u) + 1}
  {1 - \tilde\phi_\Out(u) + \tilde\chi_1(u)} \,, \\
\tilde j_{\emptyset|\In}(u) &= \tilde\phi_0(u) \frac
{\tilde\phi_{\In}^{\emptyset}(u)}
 {1 - \tilde\phi_\Out(u) + \tilde\chi_1(u)} \,.
\end{align}
\end{subequations}
The full solution of the regularized two-domain model [\cref{flrenout1L,flreninstop1L,flrenin1L,rhooutLap,rhoinLap,rhoabsLap}] is given in terms of the common denominator $\tilde\chi_2(u):=1 - \tilde\phi_\In(u) \tilde\phi_\Out(u)$:
\begin{subequations}
\label{eq:sol_extended}
\begin{align}
\tilde\rho_\In(u) &= \frac{\tilde\phi_0(u)}{u \tilde\chi_2(u)}
[1 - \tilde\phi_\In(u) - \tilde\phi_{\In}^{\emptyset}(u)] \,,
\\
\tilde\rho_\Out(u) &=
\frac{1}{u} - \frac{\tilde\phi_0(u)}{u \tilde\chi_2(u)} [1 - \tilde\phi_\In(u)] \,,
\\
\tilde\rho_{\emptyset}(u) &= \frac{\tilde\phi_0(u)}{u \tilde\chi_2(u)}
\tilde\phi_{\In}^{\emptyset}(u) \,,
\\
\tilde j_\In^-(u) &=
\frac{\tilde\phi_0(u)}{ \tilde\chi_2(u)} \tilde\phi_\In(u) \,,
\\
\tilde j_\Out^-(u) &=
\frac{\tilde\phi_0(u)}{ \tilde\chi_2(u)} \,,
\\
\tilde j_{\emptyset|\In}(u) &=
\frac{\tilde\phi_0(u)}{ \tilde\chi_2(u)} \tilde\phi_{\In}^{\emptyset}(u) \,.
\end{align}
\end{subequations}
Finally, we quote the solution to the regularized two-domain model for the case when the trajectory starts on the inner domain [\cref{joutiLap,jabsiLap,jiniLap,rhooutiLap,rhoiniLap,rhoabsiLap}]:
\begin{subequations}
\label{sec:sol_inner}
\begin{align}
	\tilde\rho_\Out(u) &= \frac{\tilde\phi_{0}^{\In}(u)}{u \tilde\chi_2(u)} [1-\tilde\phi_\Out(u)] \,,
	\\
	\tilde\rho_\In(u) &=  \frac{1 - \tilde\phi_{0}^{\emptyset}(u)}{u}
    - \frac{\tilde\phi_{0}^{\In}(u)}{u \tilde\chi_2(u)} \{1 - \tilde\phi_\Out(u) [1-\tilde\phi_{\In}^{\emptyset}(u)] \} \,,
	\\
	\tilde\rho_{\emptyset}(u) &= \frac{\tilde\phi_{0}^{\emptyset}(u)}{u} +
	\frac{\tilde\phi_{0}^{\In}(u)}{u \tilde\chi_2(u)} \, \tilde\phi_{\In}^{\emptyset}(u)\tilde\phi_\Out(u) \,,
	\\
	\tilde j_\Out^-(u) &=	\frac{\tilde\phi_{0}^{\In}(u)}{\tilde\chi_2(u)} \, \tilde\phi_\Out(u) \,,
	\\
	\tilde j_\In^-(u) &=\frac{\tilde\phi_{0}^{\In}(u)}{\tilde\chi_2(u)} \,,
	\\
	\tilde j_{\emptyset|\In}(u) &= \tilde\phi_{0}^{\emptyset}(u) +
	\frac{\tilde\phi_{0}^{\In}(u)}{\tilde\chi_2(u)} \, \tilde\phi_{\In}^{\emptyset}(u)\tilde\phi_\Out(u) \,.
	\end{align}
\end{subequations}

\bibliography{reaction_diffusion}

%Control: production of eprint (-1) disabled
\begin{thebibliography}{41}%
\makeatletter
\providecommand \@ifxundefined [1]{%
 \@ifx{#1\undefined}
}%
\providecommand \@ifnum [1]{%
 \ifnum #1\expandafter \@firstoftwo
 \else \expandafter \@secondoftwo
 \fi
}%
\providecommand \@ifx [1]{%
 \ifx #1\expandafter \@firstoftwo
 \else \expandafter \@secondoftwo
 \fi
}%
\providecommand \natexlab [1]{#1}%
\providecommand \enquote  [1]{``#1''}%
\providecommand \bibnamefont  [1]{#1}%
\providecommand \bibfnamefont [1]{#1}%
\providecommand \citenamefont [1]{#1}%
\providecommand \href@noop [0]{\@secondoftwo}%
\providecommand \href [0]{\begingroup \@sanitize@url \@href}%
\providecommand \@href[1]{\@@startlink{#1}\@@href}%
\providecommand \@@href[1]{\endgroup#1\@@endlink}%
\providecommand \@sanitize@url [0]{\catcode `\\12\catcode `\$12\catcode
  `\&12\catcode `\#12\catcode `\^12\catcode `\_12\catcode `\%12\relax}%
\providecommand \@@startlink[1]{}%
\providecommand \@@endlink[0]{}%
\providecommand \url  [0]{\begingroup\@sanitize@url \@url }%
\providecommand \@url [1]{\endgroup\@href {#1}{\urlprefix }}%
\providecommand \urlprefix  [0]{URL }%
\providecommand \Eprint [0]{\href }%
\providecommand \doibase [0]{http://dx.doi.org/}%
\providecommand \selectlanguage [0]{\@gobble}%
\providecommand \bibinfo  [0]{\@secondoftwo}%
\providecommand \bibfield  [0]{\@secondoftwo}%
\providecommand \translation [1]{[#1]}%
\providecommand \BibitemOpen [0]{}%
\providecommand \bibitemStop [0]{}%
\providecommand \bibitemNoStop [0]{.\EOS\space}%
\providecommand \EOS [0]{\spacefactor3000\relax}%
\providecommand \BibitemShut  [1]{\csname bibitem#1\endcsname}%
\let\auto@bib@innerbib\@empty
%</preamble>
\bibitem [{\citenamefont {Zhou}, \citenamefont {Rivas},\ and\ \citenamefont
  {Minton}(2008)}]{Zhou:2008}%
  \BibitemOpen
  \bibfield  {author} {\bibinfo {author} {\bibfnamefont {H.-X.}\ \bibnamefont
  {Zhou}}, \bibinfo {author} {\bibfnamefont {G.}~\bibnamefont {Rivas}}, \ and\
  \bibinfo {author} {\bibfnamefont {A.~P.}\ \bibnamefont {Minton}},\ }\bibfield
   {title} {\enquote {\bibinfo {title} {Macromolecular crowding and
  confinement: {B}iochemical, biophysical, and potential physiological
  consequences},}\ }\href {\doibase 10.1146/annurev.biophys.37.032807.125817}
  {\bibfield  {journal} {\bibinfo  {journal} {Ann. Rev. Biophys.}\ }\textbf
  {\bibinfo {volume} {37}},\ \bibinfo {pages} {375} (\bibinfo {year}
  {2008})}\BibitemShut {NoStop}%
\bibitem [{\citenamefont {Höf{}ling}\ and\ \citenamefont
  {Franosch}(2013)}]{Hoefling:2013}%
  \BibitemOpen
  \bibfield  {author} {\bibinfo {author} {\bibfnamefont {F.}~\bibnamefont
  {Höf{}ling}}\ and\ \bibinfo {author} {\bibfnamefont {T.}~\bibnamefont
  {Franosch}},\ }\bibfield  {title} {\enquote {\bibinfo {title} {Anomalous
  transport in the crowded world of biological cells},}\ }\href {\doibase
  10.1088/0034-4885/76/4/046602} {\bibfield  {journal} {\bibinfo  {journal}
  {Rep. Prog. Phys.}\ }\textbf {\bibinfo {volume} {76}},\ \bibinfo {pages}
  {046602} (\bibinfo {year} {2013})}\BibitemShut {NoStop}%
\bibitem [{\citenamefont {Weiss}(2014)}]{Weiss:2014}%
  \BibitemOpen
  \bibfield  {author} {\bibinfo {author} {\bibfnamefont {M.}~\bibnamefont
  {Weiss}},\ }\bibfield  {title} {\enquote {\bibinfo {title} {Crowding,
  diffusion, and biochemical reactions},}\ }in\ \href {\doibase
  10.1016/B978-0-12-800046-5.00011-4} {\emph {\bibinfo {booktitle} {New Models
  of the Cell Nucleus: Crowding, Entropic Forces, Phase Separation, and
  Fractals}}},\ \bibinfo {series} {Int. Rev. Cell Mol. Biol.}, Vol.\ \bibinfo
  {volume} {307},\ \bibinfo {editor} {edited by\ \bibinfo {editor}
  {\bibfnamefont {R.}~\bibnamefont {Hancock}}\ and\ \bibinfo {editor}
  {\bibfnamefont {K.~W.}\ \bibnamefont {Jeon}}}\ (\bibinfo  {publisher}
  {Academic Press},\ \bibinfo {year} {2014})\ Chap.~\bibinfo {chapter} {11},
  pp.\ \bibinfo {pages} {383--417}\BibitemShut {NoStop}%
\bibitem [{\citenamefont {Dross}\ \emph {et~al.}(2009)\citenamefont {Dross},
  \citenamefont {Spriet}, \citenamefont {Zwerger}, \citenamefont {Müller},
  \citenamefont {Waldeck},\ and\ \citenamefont {Langowski}}]{Dross:PLoS2009}%
  \BibitemOpen
  \bibfield  {author} {\bibinfo {author} {\bibfnamefont {N.}~\bibnamefont
  {Dross}}, \bibinfo {author} {\bibfnamefont {C.}~\bibnamefont {Spriet}},
  \bibinfo {author} {\bibfnamefont {M.}~\bibnamefont {Zwerger}}, \bibinfo
  {author} {\bibfnamefont {G.}~\bibnamefont {Müller}}, \bibinfo {author}
  {\bibfnamefont {W.}~\bibnamefont {Waldeck}}, \ and\ \bibinfo {author}
  {\bibfnamefont {J.}~\bibnamefont {Langowski}},\ }\bibfield  {title} {\enquote
  {\bibinfo {title} {Mapping {eGFP} oligomer mobility in living cell nuclei},}\
  }\href {\doibase 10.1371/journal.pone.0005041} {\bibfield  {journal}
  {\bibinfo  {journal} {{PLoS} {ONE}}\ }\textbf {\bibinfo {volume} {4}},\
  \bibinfo {pages} {e5041} (\bibinfo {year} {2009})}\BibitemShut {NoStop}%
\bibitem [{\citenamefont {Schlimpert}\ \emph {et~al.}(2012)\citenamefont
  {Schlimpert}, \citenamefont {Klein}, \citenamefont {Briegel}, \citenamefont
  {Hughes}, \citenamefont {Kahnt}, \citenamefont {Bolte}, \citenamefont
  {Maier}, \citenamefont {Brun}, \citenamefont {Jensen}, \citenamefont
  {Gitai},\ and\ \citenamefont {Thanbichler}}]{Schlimpert:Cell2012}%
  \BibitemOpen
  \bibfield  {author} {\bibinfo {author} {\bibfnamefont {S.}~\bibnamefont
  {Schlimpert}}, \bibinfo {author} {\bibfnamefont {E.~A.}\ \bibnamefont
  {Klein}}, \bibinfo {author} {\bibfnamefont {A.}~\bibnamefont {Briegel}},
  \bibinfo {author} {\bibfnamefont {V.}~\bibnamefont {Hughes}}, \bibinfo
  {author} {\bibfnamefont {J.}~\bibnamefont {Kahnt}}, \bibinfo {author}
  {\bibfnamefont {K.}~\bibnamefont {Bolte}}, \bibinfo {author} {\bibfnamefont
  {U.~G.}\ \bibnamefont {Maier}}, \bibinfo {author} {\bibfnamefont {Y.~V.}\
  \bibnamefont {Brun}}, \bibinfo {author} {\bibfnamefont {G.~J.}\ \bibnamefont
  {Jensen}}, \bibinfo {author} {\bibfnamefont {Z.}~\bibnamefont {Gitai}}, \
  and\ \bibinfo {author} {\bibfnamefont {M.}~\bibnamefont {Thanbichler}},\
  }\bibfield  {title} {\enquote {\bibinfo {title} {General protein diffusion
  barriers create compartments within bacterial cells},}\ }\href {\doibase
  10.1016/j.cell.2012.10.046} {\bibfield  {journal} {\bibinfo  {journal}
  {Cell}\ }\textbf {\bibinfo {volume} {151}},\ \bibinfo {pages} {1270}
  (\bibinfo {year} {2012})}\BibitemShut {NoStop}%
\bibitem [{\citenamefont {Sezgin}, \citenamefont {Davis},\ and\ \citenamefont
  {Eggeling}(2015)}]{Sezgin:Cell2015}%
  \BibitemOpen
  \bibfield  {author} {\bibinfo {author} {\bibfnamefont {E.}~\bibnamefont
  {Sezgin}}, \bibinfo {author} {\bibfnamefont {S.~J.}\ \bibnamefont {Davis}}, \
  and\ \bibinfo {author} {\bibfnamefont {C.}~\bibnamefont {Eggeling}},\
  }\bibfield  {title} {\enquote {\bibinfo {title} {Membrane
  nanoclusters{\textemdash}tails of the unexpected},}\ }\href {\doibase
  10.1016/j.cell.2015.04.008} {\bibfield  {journal} {\bibinfo  {journal}
  {Cell}\ }\textbf {\bibinfo {volume} {161}},\ \bibinfo {pages} {433} (\bibinfo
  {year} {2015})},\ \bibinfo {note} {see also references therein}\BibitemShut
  {NoStop}%
\bibitem [{\citenamefont {Raghupathy}\ \emph {et~al.}(2015)\citenamefont
  {Raghupathy}, \citenamefont {Anilkumar}, \citenamefont {Polley},
  \citenamefont {Singh}, \citenamefont {Yadav}, \citenamefont {Johnson},
  \citenamefont {Suryawanshi}, \citenamefont {Saikam}, \citenamefont {Sawant},
  \citenamefont {Panda}, \citenamefont {Guo}, \citenamefont {Vishwakarma},
  \citenamefont {Rao},\ and\ \citenamefont {Mayor}}]{Raghupathy:Cell2015}%
  \BibitemOpen
  \bibfield  {author} {\bibinfo {author} {\bibfnamefont {R.}~\bibnamefont
  {Raghupathy}}, \bibinfo {author} {\bibfnamefont {A.~A.}\ \bibnamefont
  {Anilkumar}}, \bibinfo {author} {\bibfnamefont {A.}~\bibnamefont {Polley}},
  \bibinfo {author} {\bibfnamefont {P.~P.}\ \bibnamefont {Singh}}, \bibinfo
  {author} {\bibfnamefont {M.}~\bibnamefont {Yadav}}, \bibinfo {author}
  {\bibfnamefont {C.}~\bibnamefont {Johnson}}, \bibinfo {author} {\bibfnamefont
  {S.}~\bibnamefont {Suryawanshi}}, \bibinfo {author} {\bibfnamefont
  {V.}~\bibnamefont {Saikam}}, \bibinfo {author} {\bibfnamefont {S.~D.}\
  \bibnamefont {Sawant}}, \bibinfo {author} {\bibfnamefont {A.}~\bibnamefont
  {Panda}}, \bibinfo {author} {\bibfnamefont {Z.}~\bibnamefont {Guo}}, \bibinfo
  {author} {\bibfnamefont {R.~A.}\ \bibnamefont {Vishwakarma}}, \bibinfo
  {author} {\bibfnamefont {M.}~\bibnamefont {Rao}}, \ and\ \bibinfo {author}
  {\bibfnamefont {S.}~\bibnamefont {Mayor}},\ }\bibfield  {title} {\enquote
  {\bibinfo {title} {Transbilayer lipid interactions mediate nanoclustering of
  lipid-anchored proteins},}\ }\href {\doibase 10.1016/j.cell.2015.03.048}
  {\bibfield  {journal} {\bibinfo  {journal} {Cell}\ }\textbf {\bibinfo
  {volume} {161}},\ \bibinfo {pages} {581} (\bibinfo {year}
  {2015})}\BibitemShut {NoStop}%
\bibitem [{\citenamefont {Kolds{\o}}\ \emph {et~al.}(2016)\citenamefont
  {Kolds{\o}}, \citenamefont {Reddy}, \citenamefont {Fowler}, \citenamefont
  {Duncan},\ and\ \citenamefont {Sansom}}]{Koldso:JPCB2016}%
  \BibitemOpen
  \bibfield  {author} {\bibinfo {author} {\bibfnamefont {H.}~\bibnamefont
  {Kolds{\o}}}, \bibinfo {author} {\bibfnamefont {T.}~\bibnamefont {Reddy}},
  \bibinfo {author} {\bibfnamefont {P.~W.}\ \bibnamefont {Fowler}}, \bibinfo
  {author} {\bibfnamefont {A.~L.}\ \bibnamefont {Duncan}}, \ and\ \bibinfo
  {author} {\bibfnamefont {M.~S.~P.}\ \bibnamefont {Sansom}},\ }\bibfield
  {title} {\enquote {\bibinfo {title} {Membrane compartmentalization reducing
  the mobility of lipids and proteins within a model plasma membrane},}\ }\href
  {\doibase 10.1021/acs.jpcb.6b05846} {\bibfield  {journal} {\bibinfo
  {journal} {J. Phys. Chem. B}\ }\textbf {\bibinfo {volume} {120}},\ \bibinfo
  {pages} {8873} (\bibinfo {year} {2016})}\BibitemShut {NoStop}%
\bibitem [{\citenamefont {Witzel}\ \emph {et~al.}(2019)\citenamefont {Witzel},
  \citenamefont {Götz}, \citenamefont {Lanoisel{\'{e}}e}, \citenamefont
  {Franosch}, \citenamefont {Grebenkov},\ and\ \citenamefont
  {Heinrich}}]{Witzel:BJ2019}%
  \BibitemOpen
  \bibfield  {author} {\bibinfo {author} {\bibfnamefont {P.}~\bibnamefont
  {Witzel}}, \bibinfo {author} {\bibfnamefont {M.}~\bibnamefont {Götz}},
  \bibinfo {author} {\bibfnamefont {Y.}~\bibnamefont {Lanoisel{\'{e}}e}},
  \bibinfo {author} {\bibfnamefont {T.}~\bibnamefont {Franosch}}, \bibinfo
  {author} {\bibfnamefont {D.~S.}\ \bibnamefont {Grebenkov}}, \ and\ \bibinfo
  {author} {\bibfnamefont {D.}~\bibnamefont {Heinrich}},\ }\bibfield  {title}
  {\enquote {\bibinfo {title} {Heterogeneities shape passive intracellular
  transport},}\ }\href {\doibase 10.1016/j.bpj.2019.06.009} {\bibfield
  {journal} {\bibinfo  {journal} {Biophys. J.}\ }\textbf {\bibinfo {volume}
  {117}},\ \bibinfo {pages} {203} (\bibinfo {year} {2019})}\BibitemShut
  {NoStop}%
\bibitem [{\citenamefont {Mirny}\ \emph {et~al.}(2009)\citenamefont {Mirny},
  \citenamefont {Slutsky}, \citenamefont {Wunderlich}, \citenamefont {Tafvizi},
  \citenamefont {Leith},\ and\ \citenamefont {Kosmrlj}}]{Mirny:JPA2009}%
  \BibitemOpen
  \bibfield  {author} {\bibinfo {author} {\bibfnamefont {L.}~\bibnamefont
  {Mirny}}, \bibinfo {author} {\bibfnamefont {M.}~\bibnamefont {Slutsky}},
  \bibinfo {author} {\bibfnamefont {Z.}~\bibnamefont {Wunderlich}}, \bibinfo
  {author} {\bibfnamefont {A.}~\bibnamefont {Tafvizi}}, \bibinfo {author}
  {\bibfnamefont {J.}~\bibnamefont {Leith}}, \ and\ \bibinfo {author}
  {\bibfnamefont {A.}~\bibnamefont {Kosmrlj}},\ }\bibfield  {title} {\enquote
  {\bibinfo {title} {How a protein searches for its site on {DNA}: the
  mechanism of facilitated diffusion},}\ }\href {\doibase
  10.1088/1751-8113/42/43/434013} {\bibfield  {journal} {\bibinfo  {journal}
  {J. Phys. A}\ }\textbf {\bibinfo {volume} {42}},\ \bibinfo {pages} {434013}
  (\bibinfo {year} {2009})}\BibitemShut {NoStop}%
\bibitem [{\citenamefont {Lin}, \citenamefont {Kim},\ and\ \citenamefont
  {Dzubiella}(2020)}]{Lin:2020}%
  \BibitemOpen
  \bibfield  {author} {\bibinfo {author} {\bibfnamefont {Y.-C.}\ \bibnamefont
  {Lin}}, \bibinfo {author} {\bibfnamefont {W.~K.}\ \bibnamefont {Kim}}, \ and\
  \bibinfo {author} {\bibfnamefont {J.}~\bibnamefont {Dzubiella}},\ }\href@noop
  {} {\enquote {\bibinfo {title} {Coverage fluctuations and correlations in
  nanoparticle-catalyzed diffusion-influenced bimolecular reactions},}\ }
  (\bibinfo {year} {2020})\BibitemShut {NoStop}%
\bibitem [{\citenamefont {Li}, \citenamefont {Fu},\ and\ \citenamefont
  {Su}(2012)}]{Li:2012}%
  \BibitemOpen
  \bibfield  {author} {\bibinfo {author} {\bibfnamefont {Y.}~\bibnamefont
  {Li}}, \bibinfo {author} {\bibfnamefont {Z.-Y.}\ \bibnamefont {Fu}}, \ and\
  \bibinfo {author} {\bibfnamefont {B.-L.}\ \bibnamefont {Su}},\ }\bibfield
  {title} {\enquote {\bibinfo {title} {Hierarchically structured porous
  materials for energy conversion and storage},}\ }\href {\doibase
  10.1002/adfm.201200591} {\bibfield  {journal} {\bibinfo  {journal} {Adv.
  Funct. Mater.}\ }\textbf {\bibinfo {volume} {22}},\ \bibinfo {pages} {4634}
  (\bibinfo {year} {2012})}\BibitemShut {NoStop}%
\bibitem [{\citenamefont {Han}, \citenamefont {Fu},\ and\ \citenamefont
  {Schoch}(2008)}]{Han:2008}%
  \BibitemOpen
  \bibfield  {author} {\bibinfo {author} {\bibfnamefont {J.}~\bibnamefont
  {Han}}, \bibinfo {author} {\bibfnamefont {J.}~\bibnamefont {Fu}}, \ and\
  \bibinfo {author} {\bibfnamefont {R.~B.}\ \bibnamefont {Schoch}},\ }\bibfield
   {title} {\enquote {\bibinfo {title} {Molecular sieving using nanofilters:
  Past{,} present and future},}\ }\href {\doibase 10.1039/B714128A} {\bibfield
  {journal} {\bibinfo  {journal} {Lab Chip}\ }\textbf {\bibinfo {volume} {8}},\
  \bibinfo {pages} {23} (\bibinfo {year} {2008})}\BibitemShut {NoStop}%
\bibitem [{\citenamefont {Cejka}, \citenamefont {Zikova},\ and\ \citenamefont
  {Nachtigall}(2005)}]{MolecularSieves:2005}%
  \BibitemOpen
  \bibinfo {editor} {\bibfnamefont {J.}~\bibnamefont {Cejka}}, \bibinfo
  {editor} {\bibfnamefont {N.}~\bibnamefont {Zikova}}, \ and\ \bibinfo {editor}
  {\bibfnamefont {P.}~\bibnamefont {Nachtigall}},\ eds.,\ \href
  {http://www.sciencedirect.com/science/bookseries/01672991/158/part/PB} {\emph
  {\bibinfo {title} {Molecular Sieves: From Basic Research to Industrial
  Applications}}},\ \bibinfo {series} {Studies in Surface Science and
  Catalysis}, Vol.\ \bibinfo {volume} {158, Part B}\ (\bibinfo {year} {2005})\
  \bibinfo {note} {proceedings of the 3 International Zeolite Symposium (3
  FEZE)}\BibitemShut {NoStop}%
\bibitem [{\citenamefont {Cerbelli}, \citenamefont {Giona},\ and\ \citenamefont
  {Garofalo}(2013)}]{Cerbelli:MNFl2013}%
  \BibitemOpen
  \bibfield  {author} {\bibinfo {author} {\bibfnamefont {S.}~\bibnamefont
  {Cerbelli}}, \bibinfo {author} {\bibfnamefont {M.}~\bibnamefont {Giona}}, \
  and\ \bibinfo {author} {\bibfnamefont {F.}~\bibnamefont {Garofalo}},\
  }\bibfield  {title} {\enquote {\bibinfo {title} {Quantifying dispersion of
  finite-sized particles in deterministic lateral displacement microflow
  separators through {B}renner's macrotransport paradigm},}\ }\href {\doibase
  10.1007/s10404-013-1150-8} {\bibfield  {journal} {\bibinfo  {journal}
  {Microfluid. Nanofluid.}\ }\textbf {\bibinfo {volume} {15}},\ \bibinfo
  {pages} {431} (\bibinfo {year} {2013})}\BibitemShut {NoStop}%
\bibitem [{\citenamefont {Sahimi}(1993)}]{Sahimi:1993}%
  \BibitemOpen
  \bibfield  {author} {\bibinfo {author} {\bibfnamefont {M.}~\bibnamefont
  {Sahimi}},\ }\bibfield  {title} {\enquote {\bibinfo {title} {Flow phenomena
  in rocks: from continuum models to fractals, percolation, cellular automata,
  and simulated annealing},}\ }\href {\doibase 10.1103/RevModPhys.65.1393}
  {\bibfield  {journal} {\bibinfo  {journal} {Rev. Mod. Phys.}\ }\textbf
  {\bibinfo {volume} {65}},\ \bibinfo {pages} {1393} (\bibinfo {year}
  {1993})}\BibitemShut {NoStop}%
\bibitem [{\citenamefont {Dentz}\ and\ \citenamefont
  {Castro}(2009)}]{Dentz:GeoRL09}%
  \BibitemOpen
  \bibfield  {author} {\bibinfo {author} {\bibfnamefont {M.}~\bibnamefont
  {Dentz}}\ and\ \bibinfo {author} {\bibfnamefont {A.}~\bibnamefont {Castro}},\
  }\bibfield  {title} {\enquote {\bibinfo {title} {Effective transport dynamics
  in porous media with heterogeneous retardation properties},}\ }\href
  {\doibase 10.1029/2008GL036846} {\bibfield  {journal} {\bibinfo  {journal}
  {Geophys. Res. Lett.}\ }\textbf {\bibinfo {volume} {36}},\ \bibinfo {pages}
  {L03403} (\bibinfo {year} {2009})}\BibitemShut {NoStop}%
\bibitem [{\citenamefont {Burov}(2017)}]{Bur:PRE17}%
  \BibitemOpen
  \bibfield  {author} {\bibinfo {author} {\bibfnamefont {S.}~\bibnamefont
  {Burov}},\ }\bibfield  {title} {\enquote {\bibinfo {title} {From quenched
  disorder to continuous time random walk},}\ }\href {\doibase
  10.1103/PhysRevE.96.050103} {\bibfield  {journal} {\bibinfo  {journal} {Phys.
  Rev. E}\ }\textbf {\bibinfo {volume} {96}},\ \bibinfo {pages} {050103(R)}
  (\bibinfo {year} {2017})}\BibitemShut {NoStop}%
\bibitem [{\citenamefont {Spanner}\ \emph {et~al.}(2016)\citenamefont
  {Spanner}, \citenamefont {Höf{}ling}, \citenamefont {Kapfer}, \citenamefont
  {Mecke}, \citenamefont {Schröder-Turk},\ and\ \citenamefont
  {Franosch}}]{Spanner:PRL2016}%
  \BibitemOpen
  \bibfield  {author} {\bibinfo {author} {\bibfnamefont {M.}~\bibnamefont
  {Spanner}}, \bibinfo {author} {\bibfnamefont {F.}~\bibnamefont {Höf{}ling}},
  \bibinfo {author} {\bibfnamefont {S.~C.}\ \bibnamefont {Kapfer}}, \bibinfo
  {author} {\bibfnamefont {K.~R.}\ \bibnamefont {Mecke}}, \bibinfo {author}
  {\bibfnamefont {G.~E.}\ \bibnamefont {Schröder-Turk}}, \ and\ \bibinfo
  {author} {\bibfnamefont {T.}~\bibnamefont {Franosch}},\ }\bibfield  {title}
  {\enquote {\bibinfo {title} {Splitting of the universality class of anomalous
  transport in crowded media},}\ }\href {\doibase
  10.1103/PhysRevLett.116.060601} {\bibfield  {journal} {\bibinfo  {journal}
  {Phys. Rev. Lett.}\ }\textbf {\bibinfo {volume} {116}},\ \bibinfo {pages}
  {060601} (\bibinfo {year} {2016})}\BibitemShut {NoStop}%
\bibitem [{\citenamefont {Adrover}\ \emph {et~al.}(2019)\citenamefont
  {Adrover}, \citenamefont {Passaretti}, \citenamefont {Venditti},\ and\
  \citenamefont {Giona}}]{Adrover:PhFl2019}%
  \BibitemOpen
  \bibfield  {author} {\bibinfo {author} {\bibfnamefont {A.}~\bibnamefont
  {Adrover}}, \bibinfo {author} {\bibfnamefont {C.}~\bibnamefont {Passaretti}},
  \bibinfo {author} {\bibfnamefont {C.}~\bibnamefont {Venditti}}, \ and\
  \bibinfo {author} {\bibfnamefont {M.}~\bibnamefont {Giona}},\ }\bibfield
  {title} {\enquote {\bibinfo {title} {Exact moment analysis of transient
  dispersion properties in periodic media},}\ }\href {\doibase
  10.1063/1.5127278} {\bibfield  {journal} {\bibinfo  {journal} {Phys. Fluids}\
  }\textbf {\bibinfo {volume} {31}},\ \bibinfo {pages} {112002} (\bibinfo
  {year} {2019})}\BibitemShut {NoStop}%
\bibitem [{\citenamefont {Kuzmak}\ \emph {et~al.}(2019)\citenamefont {Kuzmak},
  \citenamefont {Carmali}, \citenamefont {von Lieres}, \citenamefont
  {Russell},\ and\ \citenamefont {Kondrat}}]{Kuzmak:SR2019}%
  \BibitemOpen
  \bibfield  {author} {\bibinfo {author} {\bibfnamefont {A.}~\bibnamefont
  {Kuzmak}}, \bibinfo {author} {\bibfnamefont {S.}~\bibnamefont {Carmali}},
  \bibinfo {author} {\bibfnamefont {E.}~\bibnamefont {von Lieres}}, \bibinfo
  {author} {\bibfnamefont {A.~J.}\ \bibnamefont {Russell}}, \ and\ \bibinfo
  {author} {\bibfnamefont {S.}~\bibnamefont {Kondrat}},\ }\bibfield  {title}
  {\enquote {\bibinfo {title} {Can enzyme proximity accelerate cascade
  reactions?}}\ }\href {\doibase 10.1038/s41598-018-37034-3} {\bibfield
  {journal} {\bibinfo  {journal} {Sci. Rep.}\ }\textbf {\bibinfo {volume}
  {9}},\ \bibinfo {pages} {455} (\bibinfo {year} {2019})}\BibitemShut {NoStop}%
\bibitem [{\citenamefont {Mattos}\ \emph {et~al.}(2012)\citenamefont {Mattos},
  \citenamefont {Mej{\'{\i}}a-Monasterio}, \citenamefont {Metzler},\ and\
  \citenamefont {Oshanin}}]{Mattos:PRE2012}%
  \BibitemOpen
  \bibfield  {author} {\bibinfo {author} {\bibfnamefont {T.~G.}\ \bibnamefont
  {Mattos}}, \bibinfo {author} {\bibfnamefont {C.}~\bibnamefont
  {Mej{\'{\i}}a-Monasterio}}, \bibinfo {author} {\bibfnamefont
  {R.}~\bibnamefont {Metzler}}, \ and\ \bibinfo {author} {\bibfnamefont
  {G.}~\bibnamefont {Oshanin}},\ }\bibfield  {title} {\enquote {\bibinfo
  {title} {First passages in bounded domains: {W}hen is the mean first passage
  time meaningful?}}\ }\href {\doibase 10.1103/physreve.86.031143} {\bibfield
  {journal} {\bibinfo  {journal} {Phys. Rev. E}\ }\textbf {\bibinfo {volume}
  {86}},\ \bibinfo {pages} {031143} (\bibinfo {year} {2012})}\BibitemShut
  {NoStop}%
\bibitem [{\citenamefont {Grebenkov}, \citenamefont {Metzler},\ and\
  \citenamefont {Oshanin}(2018)}]{Grebenkov:CC2018}%
  \BibitemOpen
  \bibfield  {author} {\bibinfo {author} {\bibfnamefont {D.~S.}\ \bibnamefont
  {Grebenkov}}, \bibinfo {author} {\bibfnamefont {R.}~\bibnamefont {Metzler}},
  \ and\ \bibinfo {author} {\bibfnamefont {G.}~\bibnamefont {Oshanin}},\
  }\bibfield  {title} {\enquote {\bibinfo {title} {Strong defocusing of
  molecular reaction times results from an interplay of geometry and reaction
  control},}\ }\href {\doibase 10.1038/s42004-018-0096-x} {\bibfield  {journal}
  {\bibinfo  {journal} {Commun. Chem.}\ }\textbf {\bibinfo {volume} {1}},\
  \bibinfo {pages} {96} (\bibinfo {year} {2018})}\BibitemShut {NoStop}%
\bibitem [{\citenamefont {B{\'{e}}nichou}\ and\ \citenamefont
  {Voituriez}(2014)}]{Benichou:2014}%
  \BibitemOpen
  \bibfield  {author} {\bibinfo {author} {\bibfnamefont {O.}~\bibnamefont
  {B{\'{e}}nichou}}\ and\ \bibinfo {author} {\bibfnamefont {R.}~\bibnamefont
  {Voituriez}},\ }\bibfield  {title} {\enquote {\bibinfo {title} {From
  first-passage times of random walks in confinement to geometry-controlled
  kinetics},}\ }\href {\doibase 10.1016/j.physrep.2014.02.003} {\bibfield
  {journal} {\bibinfo  {journal} {Phys. Rep.}\ }\textbf {\bibinfo {volume}
  {539}},\ \bibinfo {pages} {225} (\bibinfo {year} {2014})}\BibitemShut
  {NoStop}%
\bibitem [{\citenamefont {Grebenkov}, \citenamefont {Metzler},\ and\
  \citenamefont {Oshanin}(2019)}]{Grebenkov:NJP2019}%
  \BibitemOpen
  \bibfield  {author} {\bibinfo {author} {\bibfnamefont {D.~S.}\ \bibnamefont
  {Grebenkov}}, \bibinfo {author} {\bibfnamefont {R.}~\bibnamefont {Metzler}},
  \ and\ \bibinfo {author} {\bibfnamefont {G.}~\bibnamefont {Oshanin}},\
  }\bibfield  {title} {\enquote {\bibinfo {title} {Full distribution of first
  exit times in the narrow escape problem},}\ }\href {\doibase
  10.1088/1367-2630/ab5de4} {\bibfield  {journal} {\bibinfo  {journal} {New J.
  Phys.}\ }\textbf {\bibinfo {volume} {21}},\ \bibinfo {pages} {122001}
  (\bibinfo {year} {2019})}\BibitemShut {NoStop}%
\bibitem [{\citenamefont {Godec}\ and\ \citenamefont
  {Metzler}(2016)}]{Godec:SR2016}%
  \BibitemOpen
  \bibfield  {author} {\bibinfo {author} {\bibfnamefont {A.}~\bibnamefont
  {Godec}}\ and\ \bibinfo {author} {\bibfnamefont {R.}~\bibnamefont
  {Metzler}},\ }\bibfield  {title} {\enquote {\bibinfo {title} {First passage
  time distribution in heterogeneity controlled kinetics: going beyond the mean
  first passage time},}\ }\href {\doibase 10.1038/srep20349} {\bibfield
  {journal} {\bibinfo  {journal} {Sci. Rep.}\ }\textbf {\bibinfo {volume}
  {6}},\ \bibinfo {pages} {20349} (\bibinfo {year} {2016})}\BibitemShut
  {NoStop}%
\bibitem [{\citenamefont {Doi}(1975)}]{Doi:1975a}%
  \BibitemOpen
  \bibfield  {author} {\bibinfo {author} {\bibfnamefont {M.}~\bibnamefont
  {Doi}},\ }\bibfield  {title} {\enquote {\bibinfo {title} {Theory of
  diffusion-controlled reactions between non-simple molecules. {I}},}\ }\href
  {\doibase 10.1016/0301-0104(75)80043-7} {\bibfield  {journal} {\bibinfo
  {journal} {Chem. Phys.}\ }\textbf {\bibinfo {volume} {11}},\ \bibinfo {pages}
  {107} (\bibinfo {year} {1975})}\BibitemShut {NoStop}%
\bibitem [{\citenamefont {Erban}\ and\ \citenamefont
  {Chapman}(2009)}]{Erban:2009}%
  \BibitemOpen
  \bibfield  {author} {\bibinfo {author} {\bibfnamefont {R.}~\bibnamefont
  {Erban}}\ and\ \bibinfo {author} {\bibfnamefont {S.~J.}\ \bibnamefont
  {Chapman}},\ }\bibfield  {title} {\enquote {\bibinfo {title} {Stochastic
  modelling of reaction–diffusion processes: algorithms for bimolecular
  reactions},}\ }\href {http://stacks.iop.org/1478-3975/6/i=4/a=046001}
  {\bibfield  {journal} {\bibinfo  {journal} {Phys. Biol.}\ }\textbf {\bibinfo
  {volume} {6}},\ \bibinfo {pages} {046001} (\bibinfo {year}
  {2009})}\BibitemShut {NoStop}%
\bibitem [{\citenamefont {Dibak}\ \emph {et~al.}(2019)\citenamefont {Dibak},
  \citenamefont {Fröhner}, \citenamefont {Noé},\ and\ \citenamefont
  {Höf{}ling}}]{Dibak:JCP2019}%
  \BibitemOpen
  \bibfield  {author} {\bibinfo {author} {\bibfnamefont {M.}~\bibnamefont
  {Dibak}}, \bibinfo {author} {\bibfnamefont {C.}~\bibnamefont {Fröhner}},
  \bibinfo {author} {\bibfnamefont {F.}~\bibnamefont {Noé}}, \ and\ \bibinfo
  {author} {\bibfnamefont {F.}~\bibnamefont {Höf{}ling}},\ }\bibfield  {title}
  {\enquote {\bibinfo {title} {Diffusion-influenced reaction rates in the
  presence of pair interactions},}\ }\href {\doibase 10.1063/1.5124728}
  {\bibfield  {journal} {\bibinfo  {journal} {J. Chem. Phys.}\ }\textbf
  {\bibinfo {volume} {151}},\ \bibinfo {pages} {164105} (\bibinfo {year}
  {2019})}\BibitemShut {NoStop}%
\bibitem [{\citenamefont {Winkelmann}\ and\ \citenamefont
  {Schütte}(2016)}]{Winkelmann:JCP2016}%
  \BibitemOpen
  \bibfield  {author} {\bibinfo {author} {\bibfnamefont {S.}~\bibnamefont
  {Winkelmann}}\ and\ \bibinfo {author} {\bibfnamefont {C.}~\bibnamefont
  {Schütte}},\ }\bibfield  {title} {\enquote {\bibinfo {title} {The
  spatiotemporal master equation: Approximation of reaction-diffusion dynamics
  via {Markov} state modeling},}\ }\href {\doibase 10.1063/1.4971163}
  {\bibfield  {journal} {\bibinfo  {journal} {J. Chem. Phys.}\ }\textbf
  {\bibinfo {volume} {145}},\ \bibinfo {pages} {214107} (\bibinfo {year}
  {2016})}\BibitemShut {NoStop}%
\bibitem [{\citenamefont {Lindenberg}\ and\ \citenamefont
  {Cukier}(1975)}]{Lindenberg:1975}%
  \BibitemOpen
  \bibfield  {author} {\bibinfo {author} {\bibfnamefont {K.}~\bibnamefont
  {Lindenberg}}\ and\ \bibinfo {author} {\bibfnamefont {R.~I.}\ \bibnamefont
  {Cukier}},\ }\bibfield  {title} {\enquote {\bibinfo {title} {Generalized
  stochastic model for molecular rotational motion in dense media},}\ }\href
  {\doibase 10.1063/1.430880} {\bibfield  {journal} {\bibinfo  {journal} {J.
  Chem. Phys.}\ }\textbf {\bibinfo {volume} {62}},\ \bibinfo {pages} {3271}
  (\bibinfo {year} {1975})}\BibitemShut {NoStop}%
\bibitem [{\citenamefont {Weiss}(1976)}]{Weiss:1976}%
  \BibitemOpen
  \bibfield  {author} {\bibinfo {author} {\bibfnamefont {G.~H.}\ \bibnamefont
  {Weiss}},\ }\bibfield  {title} {\enquote {\bibinfo {title} {The two-state
  random walk},}\ }\href {\doibase 10.1007/bf01012035} {\bibfield  {journal}
  {\bibinfo  {journal} {J. Stat. Phys.}\ }\textbf {\bibinfo {volume} {15}},\
  \bibinfo {pages} {157} (\bibinfo {year} {1976})}\BibitemShut {NoStop}%
\bibitem [{\citenamefont {Margolin}\ \emph {et~al.}(2006)\citenamefont
  {Margolin}, \citenamefont {Protasenko}, \citenamefont {Kuno},\ and\
  \citenamefont {Barkai}}]{Margolin:JPCB2006}%
  \BibitemOpen
  \bibfield  {author} {\bibinfo {author} {\bibfnamefont {G.}~\bibnamefont
  {Margolin}}, \bibinfo {author} {\bibfnamefont {V.}~\bibnamefont
  {Protasenko}}, \bibinfo {author} {\bibfnamefont {M.}~\bibnamefont {Kuno}}, \
  and\ \bibinfo {author} {\bibfnamefont {E.}~\bibnamefont {Barkai}},\
  }\bibfield  {title} {\enquote {\bibinfo {title} {Photon counting statistics
  for blinking {CdSe}-{ZnS} quantum dots: A {L}évy walk process},}\ }\href
  {\doibase 10.1021/jp061487m} {\bibfield  {journal} {\bibinfo  {journal} {J.
  Phys. Chem. B}\ }\textbf {\bibinfo {volume} {110}},\ \bibinfo {pages} {19053}
  (\bibinfo {year} {2006})}\BibitemShut {NoStop}%
\bibitem [{\citenamefont {Feller}(1968)}]{Feller:ProbabilityBd2}%
  \BibitemOpen
  \bibfield  {author} {\bibinfo {author} {\bibfnamefont {W.}~\bibnamefont
  {Feller}},\ }\href@noop {} {\emph {\bibinfo {title} {An Introduction to
  Probability Theory and Its Applications}}},\ \bibinfo {edition} {3rd}\ ed.,\
  Vol.~\bibinfo {volume} {2}\ (\bibinfo  {publisher} {Wiley},\ \bibinfo {year}
  {1968})\BibitemShut {NoStop}%
\bibitem [{\citenamefont {Hughes}(1995)}]{Hughes:Random_Walks}%
  \BibitemOpen
  \bibfield  {author} {\bibinfo {author} {\bibfnamefont {B.~D.}\ \bibnamefont
  {Hughes}},\ }\href@noop {} {\emph {\bibinfo {title} {Random Walks and Random
  Environments}}},\ Vol.\ \bibinfo {volume} {1: Random Walks}\ (\bibinfo
  {publisher} {Oxford},\ \bibinfo {address} {Clarendon Press},\ \bibinfo {year}
  {1995})\BibitemShut {NoStop}%
\bibitem [{\citenamefont {Chechkin}, \citenamefont {Gorenflo},\ and\
  \citenamefont {Sokolov}(2005)}]{Chechkin:JPA2005}%
  \BibitemOpen
  \bibfield  {author} {\bibinfo {author} {\bibfnamefont {A.~V.}\ \bibnamefont
  {Chechkin}}, \bibinfo {author} {\bibfnamefont {R.}~\bibnamefont {Gorenflo}},
  \ and\ \bibinfo {author} {\bibfnamefont {I.~M.}\ \bibnamefont {Sokolov}},\
  }\bibfield  {title} {\enquote {\bibinfo {title} {Fractional diffusion in
  inhomogeneous media},}\ }\href {\doibase 10.1088/0305-4470/38/42/l03}
  {\bibfield  {journal} {\bibinfo  {journal} {J. Phys. A}\ }\textbf {\bibinfo
  {volume} {38}},\ \bibinfo {pages} {L679} (\bibinfo {year}
  {2005})}\BibitemShut {NoStop}%
\bibitem [{\citenamefont {Froemberg}\ and\ \citenamefont
  {Sokolov}(2008)}]{Froemberg:PRL2008}%
  \BibitemOpen
  \bibfield  {author} {\bibinfo {author} {\bibfnamefont {D.}~\bibnamefont
  {Froemberg}}\ and\ \bibinfo {author} {\bibfnamefont {I.~M.}\ \bibnamefont
  {Sokolov}},\ }\bibfield  {title} {\enquote {\bibinfo {title} {Stationary
  fronts in an $\text{A}+\text{B} \to 0$ reaction under subdiffusion},}\ }\href
  {\doibase 10.1103/PhysRevLett.100.108304} {\bibfield  {journal} {\bibinfo
  {journal} {Phys. Rev. Lett.}\ }\textbf {\bibinfo {volume} {100}},\ \bibinfo
  {pages} {108304} (\bibinfo {year} {2008})}\BibitemShut {NoStop}%
\bibitem [{\citenamefont {Froemberg}\ \emph {et~al.}(2011)\citenamefont
  {Froemberg}, \citenamefont {Schmidt-Martens}, \citenamefont {Sokolov},\ and\
  \citenamefont {Sagu\'es}}]{Froemberg:PRE2011}%
  \BibitemOpen
  \bibfield  {author} {\bibinfo {author} {\bibfnamefont {D.}~\bibnamefont
  {Froemberg}}, \bibinfo {author} {\bibfnamefont {H.}~\bibnamefont
  {Schmidt-Martens}}, \bibinfo {author} {\bibfnamefont {I.~M.}\ \bibnamefont
  {Sokolov}}, \ and\ \bibinfo {author} {\bibfnamefont {F.}~\bibnamefont
  {Sagu\'es}},\ }\bibfield  {title} {\enquote {\bibinfo {title} {Asymptotic
  front behavior in an $\text{A} + \text{B} \to 2\text{A}$ reaction under
  subdiffusion},}\ }\href {\doibase 10.1103/PhysRevE.83.031101} {\bibfield
  {journal} {\bibinfo  {journal} {Phys. Rev. E}\ }\textbf {\bibinfo {volume}
  {83}},\ \bibinfo {pages} {031101} (\bibinfo {year} {2011})}\BibitemShut
  {NoStop}%
\bibitem [{\citenamefont {Straube}\ \emph {et~al.}(2020)\citenamefont
  {Straube}, \citenamefont {Kowalik}, \citenamefont {Netz},\ and\ \citenamefont
  {Höf{}ling}}]{Straube:CP2020}%
  \BibitemOpen
  \bibfield  {author} {\bibinfo {author} {\bibfnamefont {A.~V.}\ \bibnamefont
  {Straube}}, \bibinfo {author} {\bibfnamefont {B.~G.}\ \bibnamefont
  {Kowalik}}, \bibinfo {author} {\bibfnamefont {R.~R.}\ \bibnamefont {Netz}}, \
  and\ \bibinfo {author} {\bibfnamefont {F.}~\bibnamefont {Höf{}ling}},\
  }\bibfield  {title} {\enquote {\bibinfo {title} {Rapid onset of molecular
  friction in liquids bridging between the atomistic and hydrodynamic
  pictures},}\ }\href {\doibase 10.1038/s42005-020-0389-0} {\bibfield
  {journal} {\bibinfo  {journal} {Commun. Phys.}\ }\textbf {\bibinfo {volume}
  {3}},\ \bibinfo {pages} {126} (\bibinfo {year} {2020})}\BibitemShut {NoStop}%
\bibitem [{\citenamefont {Redner}(2001)}]{Redner:FirstPassage}%
  \BibitemOpen
  \bibfield  {author} {\bibinfo {author} {\bibfnamefont {S.}~\bibnamefont
  {Redner}},\ }\href@noop {} {\emph {\bibinfo {title} {A Guide to First Passage
  Processes}}}\ (\bibinfo  {publisher} {Cambridge University Press},\ \bibinfo
  {address} {New York},\ \bibinfo {year} {2001})\BibitemShut {NoStop}%
\bibitem [{\citenamefont {Grebenkov}, \citenamefont {Metzler},\ and\
  \citenamefont {Oshanin}(2020)}]{Grebenkov:NJP2020}%
  \BibitemOpen
  \bibfield  {author} {\bibinfo {author} {\bibfnamefont {D.~S.}\ \bibnamefont
  {Grebenkov}}, \bibinfo {author} {\bibfnamefont {R.}~\bibnamefont {Metzler}},
  \ and\ \bibinfo {author} {\bibfnamefont {G.}~\bibnamefont {Oshanin}},\
  }\bibfield  {title} {\enquote {\bibinfo {title} {From single-particle
  stochastic kinetics to macroscopic reaction rates: fastest first-passage time
  of $n$ random walkers},}\ }\href {\doibase 10.1088/1367-2630/abb1de}
  {\bibfield  {journal} {\bibinfo  {journal} {New J. Phys.}\ }\textbf {\bibinfo
  {volume} {22}},\ \bibinfo {pages} {103004} (\bibinfo {year}
  {2020})}\BibitemShut {NoStop}%
\end{thebibliography}%

\end{document}